\documentclass[ALICE,manyauthors]{cernphprep}

\usepackage[comma,square,numbers,sort&compress]{natbib}
\usepackage{hyperref}
\usepackage{lineno}
\pdfoutput=1
\usepackage{wasysym}
\usepackage{booktabs}
\usepackage{multirow}
\usepackage{color}

\usepackage[utf8x]{inputenc}
\usepackage{csquotes}
\usepackage{comment}
\usepackage[T1]{fontenc}

\begin{document}%

\pdfoutput=1
\begin{titlepage}
\PHyear{2019}
\PHnumber{205}      
\PHdate{26 September}  
%

\title{Measurement of electrons from semileptonic heavy-flavour hadron decays 
at midrapidity in pp 
and Pb--Pb collisions at $\boldsymbol{\sqrt{s_{\mathrm{NN}}}=5.02}$ \textbf{ TeV}}
\ShortTitle{Measurement of electrons from semileptonic heavy-flavour hadron decays}   
\Collaboration{ALICE Collaboration\thanks{See Appendix~\ref{app:collab} for the list of collaboration members}}
\ShortAuthor{ALICE Collaboration} 

\begin{abstract}
The differential invariant yield as a function of transverse momentum ($p_\mathrm{T}$) of electrons from semileptonic heavy-flavour hadron decays was measured at midrapidity in central (0--10\%), semi-central (30--50\%) and peripheral (60--80\%) lead--lead (Pb--Pb) collisions at $\sqrt{s_{\mathrm{NN}}}=5.02\text{ TeV}$ in the $p_{\mathrm{T}}$ intervals 0.5--26 GeV/$c$ (0--10\% and 30--50\%) and 0.5--10 GeV/$c$ (60--80\%). The production cross section in proton--proton (pp) collisions at $\sqrt{s}=5.02$ TeV was measured as well in $0.5<p_\mathrm{T}<10$~GeV/$c$ and it lies close to the upper band of perturbative QCD calculation uncertainties up to $p_\mathrm{T}=5$~GeV/$c$ and close to the mean value for larger $p_\mathrm{T}$.
The modification of the electron yield with respect to what is expected for an incoherent superposition of nucleon--nucleon collisions is evaluated by measuring the nuclear modification factor $R_{\mathrm{AA}}$.
The measurement of the $R_{\mathrm{AA}}$ in different centrality classes allows in-medium energy loss of charm and beauty quarks to be investigated. The $R_{\mathrm{AA}}$ shows a suppression with respect to unity at intermediate $p_\mathrm{T}$, which increases while moving towards more central collisions.
Moreover, the measured $R_{\mathrm{AA}}$ is sensitive to  the modification of the parton distribution functions (PDF) in nuclei, like nuclear shadowing, which causes a suppression of the heavy-quark production at low $p_\mathrm{T}$ in heavy-ion collisions at LHC.
\end{abstract}
\end{titlepage}
\setcounter{page}{2}

\section{Introduction}

The main goal of ALICE is the study of the Quark-Gluon Plasma (QGP), a state of matter which is expected to be created in ultra-relativistic heavy-ion collisions where high temperatures and high energy densities are reached at the LHC
\cite{bib::Alice_performance}. Due to their large masses ($m_\text{c}\approx 1.5 \mbox{ GeV/}c^2$, $m_\text{b}\approx 4.8 \mbox{ GeV/}c^2$), 
charm and beauty quarks (heavy-flavour) are mostly produced via partonic scattering processes with high momentum transfer, which have typical time scales smaller than the QGP thermalisation time (1 fm/$c$ \cite{bib::intro_thermalization}). Furthermore, additional thermal production, as well as annihilation rates, of charm and beauty quarks in the strongly-interacting matter are expected to be small in Pb--Pb collisions even at LHC energies \cite{bib::Averbeck, bib::Braun_Munziger}. Consequently, charm and beauty quarks experience the full evolution of the hot and dense medium produced in high-energy heavy-ion collisions, therefore they are ideal probes to investigate the properties of the QGP.

Quarks and gluons interact strongly with the medium and they are expected to lose energy through elastic collisions \cite{Colla,intro3} and radiative processes \cite{intro1, bib::intro2}.
Quarks have a smaller colour coupling factor with respect to gluons, hence the energy loss for quarks is expected to be smaller than that for gluons. In addition, the dead-cone effect is expected to reduce small-angle gluon radiation for heavy quarks with moderate energy to mass ratio \cite{bib::dead_cone_effect}, thus further
attenuating the effect of the medium. 
The combination of all these effects results in the observed hierarchical mass dependent energy loss \cite{bib::intro2, bib::intro_massdep_Eloss1, bib::intro_massdep_Eloss2, Zhang:2003wk, bib::intro4, Adam:2016wyz, Adam:2015nna, Acharya:2018hre, Sirunyan:2017oug
, Adam:2015rba, Sirunyan:2017isk, Sirunyan:2018ktu, TheATLAScollaboration, Andronic:2015wma}.

In order to quantify medium effects on heavy-flavour observables measured in heavy-ion collisions, they are compared with measurements in proton--proton (pp) collisions, where these effects are expected to be absent. 

In pp collisions, heavy-quark
production can be described by perturbative Quantum Chromodynamics (pQCD) calculations for all transverse momenta,
whereas pQCD is not applicable for the calculation of light quark and gluon production at low transverse momenta
\cite{bib::Averbeck}.
Moreover, measurements of heavy-flavour production cross sections in pp collisions 
provide the necessary experimental reference for heavy-ion collisions.

The medium effects on
heavy quarks are quantified through the measurement of the nuclear modification factor, defined as the ratio between the yield of particles produced in ion--ion collisions ($\mathrm{d}^2N_{\mathrm{AA}}/\mathrm{d}p_{\mathrm{T}}\mathrm{d}y$) and the cross section measured in proton-proton collisions at the same energy ($\mathrm{d}^2\sigma_{\mathrm{pp}}/\mathrm{d}p_{\mathrm{T}}\mathrm{d}y$), normalised by the average nuclear overlap function $\langle T_{\mathrm{AA}}\rangle$:
\begin{equation}
    R_{\mathrm{AA}}(p_\mathrm{T},y)=\frac{1}{\langle T_{\mathrm{AA}}\rangle}\cdot\frac{\mathrm{d}^2N_{\mathrm{AA}}/\mathrm{d}p_{\mathrm{T}}\mathrm{d}y}{\mathrm{d}^2\sigma_{\mathrm{pp}}/\mathrm{d}p_{\mathrm{T}}\mathrm{d}y}\hspace{1mm}.
    \label{eq::RAA}
\end{equation}
The $\langle T_{\text{AA}}\rangle$ is defined as the average number of nucleon--nucleon collisions $\langle N_{\textrm{coll}}\rangle$, which can be estimated via Glauber model calculations \cite{Glauber:1970jm, Miller:2007ri}, divided by the inelastic nucleon-nucleon cross section.  
In-medium energy loss shifts the transverse momenta towards lower values, therefore at intermediate and high $p_\mathrm{T}$ ($p_\mathrm{T}\apprge 2 \text{ GeV/}c$) a suppression of the production is expected ($R_{\mathrm{AA}}<1$). 
Assuming the total cross section evaluated using $\langle N_{\textrm{coll}}\rangle$ scaling is not modified, the nuclear modification factor is expected to increase towards lower $p_\mathrm{T}$, compensating the depletion at higher momenta.
Such a rise was measured by the PHENIX and STAR experiments at RHIC in Au--Au and Cu--Cu collisions at $\sqrt{s_\mathrm{NN}}=200\text{ GeV}$ for electrons from heavy-flavour hadron decays \cite{Abelev:2006db, Adare:2010de, Adamczyk:2014uip}. The nuclear modification factor for electrons from semileptonic heavy-flavour hadron decays was also measured by the ALICE collaboration in Pb--Pb collisions at $\sqrt{s_\mathrm{NN}}=2.76\text{ TeV}$ \cite{Acharya:2018upq, Adam:2016khe}, where the mentioned trend of $R_{\mathrm{AA}}$ was also observed. At low $p_\mathrm{T}$, the nuclear modification factor reaches a maximum around $1\text{ GeV/}c$ and tends to decrease at lower $p_\mathrm{T}$. This trend can be explained by initial and final state effects, like the collective expansion of the hot and dense system \cite{Adam:2016ssk, Abelev:2013lca, vanHees:2005wb}, the interplay between hadronisation via fragmentation and coalescence \cite{Greco:2003vf, Andronic:2003zv, Andronic:2015wma} and the modification of the parton distribution functions (PDF) inside bound nucleons \cite{Eskola:2009uj}. 

Initial-state effects at the LHC are explored with proton--nucleus collisions, where an extended QGP phase is not expected to be formed. The nuclear modification factor of electrons from charm and beauty hadron decays \cite{Adam:2016wyz, Adam:2015qda} and of \textit{D} mesons \cite{Abelev:2014hha} in p--Pb collisions at
$\sqrt{s_\mathrm{NN}}=5.02\text{ TeV}$  was found to be consistent with unity within uncertainties.
From this, one can conclude that the strong suppression observed in Pb--Pb collisions is due to substantial final-state interactions of heavy quarks with the QGP formed in these collisions.
However, it is important to note that recently the measurement  of the elliptic flow of electrons from semileptonic heavy-flavour hadron decays \cite{Acharya:2018dxy} and of \textit{D} mesons \cite{Sirunyan:2018toe} have been published, showing intriguing and not yet fully understood collective effects in high-multiplicity p--Pb collisions in the heavy-flavour sector.

This paper reports the measurement of the production cross section in pp collisions, the invariant yields and the nuclear modification factor, $R_\mathrm{AA}$, in Pb--Pb collisions as a function of $p_\mathrm{T}$ of electrons from semileptonic heavy-flavour hadron decays at mid-rapidity 
at the centre-of-mass energy per nucleon pair $\sqrt{s_\mathrm{NN}}=5.02\text{ TeV}$. 
In order to study how the yield and $R_\mathrm{AA}$ change with centrality in Pb--Pb collisions, the measurement was done in three representative classes: the 0-10\% class for central Pb--Pb collisions, the 30-50\% for semi-central Pb--Pb collisions and 60-80\% for peripheral Pb--Pb collisions.
\section{Experimental apparatus and data sample}
The ALICE detector is described in detail in Refs. \cite{bib::Alice_performance,Aamodt:2008zz}. The experiment mainly consists of a central barrel at midrapidity ($|\eta|<0.9$), embedded in a cylindrical solenoid which provides a magnetic field of 0.5 T parallel to the beam direction, and a muon spectrometer at forward rapidity ($-4<\eta<-2.5$).

Charged particles produced in the collisions and originating from particle decays are tracked by the Inner Tracking System (ITS) \cite{1748-0221-5-03-P03003} and the Time Projection Chamber (TPC) \cite{Alme:2010ke}. 
The ITS detector, composed of the Silicon Pixel Detector (SPD), Silicon Drift Detector (SDD), and Silicon Strip Detector (SSD), consists of six cylindrical silicon layers
surrounding the beam vacuum pipe.
These provide measurements of particle momenta and energy loss (d$E$/d$x$) used for charged-particle identification (PID), together with the TPC. The particle identification is complemented by a Time-Of-Flight (TOF) \cite{Carnesecchi:2018oss} detector, which measures the time-of-flight of charged particles.
The TOF detector distinguishes electrons from kaons, protons, and pions up to $p_\mathrm{T}\simeq 2.5$ GeV/$c$, $p_\mathrm{T}\simeq 4$ GeV/$c$ and $p_\mathrm{T}\simeq1$ GeV/$c$, respectively. 
The ElectroMagnetic Calorimeter (EMCal) \cite{Abeysekara:2010ze} 
covers a pseudorapidity region of $|\eta|<0.7$ and it is used to measure electrons, photons, and jets in an azimuthal region of $\sim107^\text{o}$.
The electron identification in the EMCal is based on the measurement of the $E/p$ ratio, where $E$ is the energy of the EMCal cluster matched to the prolongation of the track
with momentum $p$ reconstructed with the TPC and ITS detectors.
The V0 detectors \cite{vZero} consist of two arrays of 32 scintillator tiles covering the pseudorapidity ranges $2.8<\eta<5.1$ (V0A) and $-3.7<\eta<-1.7$ (V0C), respectively, and are used for event characterisation. 

The results presented in this paper are based on data samples of Pb--Pb collisions recorded in 2015 and of pp collisions at the same energy recorded in 2017.
The analysed events were collected with a minimum bias (MB) trigger 
of
a logic AND between the V0A and V0C detectors.
Pb–Pb collisions were also recorded using the EMCal trigger, which requires an EMCal cluster energy summed over a group of 4$\times$4 calorimeter cells 
larger than an energy threshold of 10 GeV. The EMCAL triggered events were used for electron measurements for $p_\mathrm{T}>12$ GeV/$c$.
The centrality classes were defined in terms of percentiles of the hadronic Pb–Pb cross section, defined by selections on the sum of the V0 signal amplitudes \cite{ALICE-PUBLIC-2018-011}.

For both collision systems,
only events with at least two tracks and a reconstructed primary vertex located between $\pm$ 10 cm with respect to the nominal interaction point along the $z$-axis are considered. 
Events affected by pile-up from different bunch
crossings, which constitute less than 1\% of the recorded sample, were rejected \cite{Acharya:2018upq}. The number of events analysed in the two collision systems with the different trigger configurations is summarised in Table \ref{tab::num_evs}, together with the average nuclear overlap function $\langle T_\mathrm{AA}\rangle$ \cite{ALICE-PUBLIC-2018-011,Loizides:2017ack}.

\begin{table}[ht!]
	\centering
	\caption{
	Number of events and $\langle T_\mathrm{AA}\rangle$ \cite{ALICE-PUBLIC-2018-011,Loizides:2017ack} used in the analysis, split by collisions system, trigger configuration, and centrality class.
	}
	\begin{tabular}{lcccc}
		\toprule
			& centrality& MB	&	EMCal trigger & $\langle T_\mathrm{AA}\rangle$ (mb$^{-1}$)\\
		\midrule
		pp	& -- &	881$\times 10^{6}$	&	--  & --\\
		\midrule
		\multirow{3}*{Pb--Pb}
		& 0--10\%       & 6 $\times 10^{6}$ & 1.2 $\times 10^{6}$ & $23.26\pm0.17$ \\
		& 30--50\%& 12 $\times 10^{6}$ & 0.3 $\times 10^{6}$ & $3.917\pm0.065$\\
		& 60--80\%&     12 $\times 10^{6}$ & -- & $0.4188\pm0.0106$\\
		\bottomrule	
	\end{tabular}
	\label{tab::num_evs}
\end{table}

\section{Data analysis}
The $p_{\rm T}$-differential yield of electrons from semileptonic heavy-flavour hadron decays is computed by 
measuring the inclusive electron yield and subtracting the contribution of electrons that do not originate 
from semileptonic heavy-flavour hadron decays.
In the following, the inclusive electron identification strategy and the subtraction of electrons 
originating from background sources are described. 

\subsection{Track selection and electron identification}

The selection criteria are similar to the ones described in Refs. \cite{Acharya:2018upq,Adam:2016khe}. They are summarised together with the kinematic cuts applied in the analyses in Table \ref{tab::track_cuts}.

It is important to note that only tracks that have hits on both SPD layers are accepted so that electrons from late photon conversions in the detector material are significantly reduced. In the Pb--Pb analysis for $p_\mathrm{T}$ $>$3 GeV/$c$, also tracks with a single hit in the SPD are considered, since the amount of photonic background starts to become negligible. 
In the analysis in which the EMCal detector is used, specific track-cluster matching criteria are adopted.

\begin{table}[ht!]
	\centering
	\caption{Track selection criteria used in the analyses.\enquote{DCA} is an abbreviation for \enquote{distance of closest approach} of a track to the primary vertex.
	}
	\begin{tabular}{cccc}
		\toprule
		Parameter	&	pp	& Pb--Pb  & Pb--Pb  \\
			&		&($p_\mathrm{T}$ $<$ 3 GeV/$c$) &  ($p_\mathrm{T}$ $>$ 3 GeV/$c$) \\
		\midrule
		$|y|$ & $<$ 0.8 & $<$ 0.8 & $<$ 0.6 \\
		Number of clusters in TPC & $\geq 100$ & $\geq 120$ & $\geq 80$ \\
	    TPC clusters  in d$E$/d$x$ calculation & $\geq 80$ & $\geq 80$ & -- \\
		Number of clusters in ITS & $\geq3$ & $\geq4$ & $\geq3$ \\
		Minimum number of clusters in SPD & 2 & 2 & 1 \\
		$|$DCA$_\text{xy}|$ & $<1$ cm & $<1$ cm & $<2.4$ cm \\
		$|$DCA$_\text{z}|$ & $<2$ cm & $<2$ cm & $<3.2$ cm \\
		Found / findable clusters in TPC & $>0.6$ & $>0.6$ & $>0.6$ \\
		$\chi^2$/clusters in TPC & $<$ 4 & $<$ 4 & $<$ 4 \\
		track-cluster matching in EMCal & -- & -- & $\sqrt{\Delta \varphi^2+\Delta \eta^2}<0.02$ \\
		\bottomrule		
	\end{tabular}
	\label{tab::track_cuts}
\end{table}

As in the procedure followed in Refs. \cite{Acharya:2018upq,Adam:2016khe}, electron candidates are identified according to the criteria listed in Table \ref{tab::PID_criteria}. These requirements depend on the data sample and on the transverse momentum interval in which the analyses are performed.

The electron identification in pp collisions is performed by evaluating the signal from the TPC and TOF detectors. The discriminant variable in the former detector is the deviation of d$E$/d$x$ from the parameterised electron Bethe-Bloch \cite{BetheBloch} expectation value, expressed in units of the d$E$/d$x$ resolution, $n_{\sigma\textrm{,e}}^{\rm TPC}$, while in the latter one the analogous variable $n_{\sigma\textrm{,e}}^{\rm TOF}$, referring to the particle time-of-flight, is considered. 
The criterion $|n_{\sigma\textrm{,e}}^\textrm{TOF}|<3$, used for electron identification up to $p_\mathrm{T}=3$ GeV/$c$, is required to reduce background from kaons and protons.
A momentum dependent criterion on $n_{\sigma\textrm{,e}}^\textrm{TPC}$ is adopted to guarantee a constant electron identification efficiency of 70\% for $p_\mathrm{T}<3$ GeV/$c$ and of 50\% for higher transverse momenta by reducing the selection window in $n_{\sigma\textrm{,e}}^\textrm{TPC}$, in order to keep the hadron contamination 
sufficiently low. In the Pb--Pb analysis for $p_\mathrm{T}<3$ GeV/$c$, the electron identification is performed by applying the same requirement on TOF and due to the large densities of tracks, a selection between $-4<n_{\sigma\textrm{,e}}^{\rm ITS}<2$ on the energy deposited in the SDD and SSD detectors is applied in all centrality classes. Finally, the selection on $n_{\sigma\textrm{,e}}^\textrm{TPC}$ ensures a constant electron identification efficiency of 50\% for all centrality classes. The hadron contamination fraction after the PID is estimated by fitting the $n_{\sigma\textrm{,e}}^{\text{TPC}}$ distribution for each particle species with an analytic function in different momentum intervals \cite{Acharya:2018upq, Adam:2016khe}.
The inclusive electron sample is then selected by applying a further criterion on $n_{\sigma\textrm{,e}}^{\text{TPC}}$, which is chosen in order to have a constant efficiency as a function of the momentum, as well as to have the hadron contamination under control. This criterion is loosened for  $p_\mathrm{T}>3$ GeV/$c$, due to the lower amount of selected hadrons when the EMCal detector is employed.

In the Pb--Pb analysis for $p_\mathrm{T}>3$ GeV/$c$, the electron candidates are first selected by the measurement of the TPC d$E$/d$x$ with the criterion $-1<n_{\sigma\textrm{,e}}^{\textrm{TPC}}<$3. Then, the selection $0.8<E/p<1.3$ on the energy over momentum ratio is applied. 
Unlike for hadrons, the ratio $E/p$ is close to 1 for electrons because they deposit most of their energy in the EMCal. 
Furthermore, the electromagnetic showers of electrons are more circular than the ones produced by hadrons. 
Generally, the shower shape produced in the calorimeter has an elliptical shape which can be characterised by its two axes: $\sigma_l^2$ for the long, and $\sigma_s^2$ for the short axis. A rather lose selection of 
$0.01<\sigma_s^2<0.35$
is chosen, since it reduces the hadron contamination while at the same time it does not affect significantly the electron signal \cite{Adam:2016khe}.
The residual hadron background in the electron sample is evaluated using the $E/p$ distribution for hadron-dominated
tracks selected with $n_{\sigma\textrm{,e}}^{\text{TPC}}<-3.5$. The $E/p$ distribution of the hadrons is then normalised to match
the distribution of the electron candidates in $0.4<E/p<0.7$ (away from the true electron peak), so that the fraction of contaminating hadrons under the electron peak can be estimated. 

\begin{table}[t]
	\centering
	\caption{Electron identification criteria. 
		The following momentum-dependent function is used for the electron identification in pp collisions, based on the TPC d$E$/d$x$: $f(p)=\textrm{Min}(0.12,0.02+0.07p)$.
		For the electron selection based on clusters in the EMCAL, a criterion on the ``$\sigma_{s}^{2}$" parameter \cite{Adam:2016khe}, corresponding to the shorter-axis of the shower shape, is used. For brevity, the ``low $p_\mathrm{T}$" label is used in place of ``$p_\mathrm{T}<3$ GeV/$c$", as well as ``high $p_\mathrm{T}$" in place of ``$p_\mathrm{T}>3$ GeV/$c$".
	}
	\begin{tabular}{lcccccc}
		\toprule
		& centrality &	n$_{\sigma\textrm{,e}}^\text{TPC}$	& n$_{\sigma\textrm{,e}}^{\text{TOF}}$ & n$_{\sigma\textrm{,e}}^{\text{ITS}}$ & $E/p$ & shower shape \\
		\midrule
		pp (low $p_\mathrm{T}$)
		& -- & $[-0.5+f(p),$ $3]$ & $[-3, 3]$ & -- & -- & -- \\
		\midrule
		pp (high $p_\mathrm{T}$) 
		& -- & $[0.12,$ $3]$ & -- & -- & -- & -- \\
		\midrule
		\multirow{3}*{Pb--Pb (low $p_\mathrm{T}$)
		}
		& 0--10\% & $[-0.16, 3]$ &  & & &  \\
		& 30--50\%& $[0, 3]$ & $[-3, 3]$ & $[-4, 2]$ & -- & --\\
		& 60--80\%& $[0.2, 3]$  &  & & &  \\
		\midrule
		\multirow{3}*{Pb--Pb (high $p_\mathrm{T}$)
		}
		& 0--10\% &  &  &  & & \\
		& 30--50\%& $[-1, 3]$ & -- & -- & [0.8, 1.3] & $0.01<\text{$\sigma_{s}^{2}$}<0.35$\\
		& 60--80\%& &  &  & & \\
		\bottomrule
	\end{tabular}
	\label{tab::PID_criteria}
\end{table}

In pp events, the hadron contamination is below $1\%$ at low $p_\mathrm{T}$, while it reaches about $40\%$ at $p_\mathrm{T}=10$~GeV/$c$. In Pb--Pb, the largest hadron contamination is measured in the most central collisions, where a contamination of about $7\%$ and $10\%$ mainly due to kaon and proton crossing the electron band at $p_\mathrm{T}=0.5$~ GeV/$c$ and $p_\mathrm{T}=1$~GeV/$c$ respectively is present. 
The total hadron contamination contribution amounts to $5\%$ at $p_\mathrm{T}=3$~GeV/$c$ in central events and tends to decrease towards more peripheral collisions.
In the EMCal analysis a maximum residual contamination of about $ 10\%$ is subtracted at the highest transverse momenta in the 0--10\% centrality class. 
In both collision systems, the hadron contamination is subtracted statistically from the inclusive electron candidate yield.

In Pb--Pb collisions, the rapidity ranges used in the ITS-TPC-TOF ($p_\mathrm{T}$ $<$ 3 GeV/$c$) and TPC-EMCal ($p_\mathrm{T}$ $>$ 3 GeV/$c$) analyses are restricted to $|y|$ $<$ 0.8 and $|y|$ $<$ 0.6, respectively, to avoid the edges of the detectors, where the systematic uncertainties related to particle identification increase.

\subsection{Subtraction of electrons from non heavy-flavour sources}
The selected inclusive electron sample does not only contain electrons from open heavy-flavour hadron decays, but also different sources of background:
\begin{enumerate}
	\item  
	electrons from Dalitz decays of light neutral mesons, mainly $\pi^0$ and $\eta$, and from photon conversions in the detector material as well as from thermal and hard scattering processes, called photonic in the following ;
	\item electrons from weak decays of kaons: 
	${\rm K^{0/\pm}} \rightarrow {\rm e}^{\pm}\pi^{\mp/0}\,\overset{\scriptscriptstyle(-)}{\nu_{e}}$ (${\rm K}_{{\rm e}3}$) ; 
	\item di-electron decays of quarkonia: J$/\psi$, $\Upsilon\to $ e$^+$e$^-$ ;
	\item di-electron decays of light vector mesons: $\omega,\phi,\rho_0\to$ e$^+$e$^-$ ;
	\item electrons from $\mathrm{W}$ and $\mathrm{Z}/\gamma^{*}$ .
\end{enumerate}
The photonic tagging method \cite{Acharya:2018upq,Adam:2016khe,Adam:2015qda, Abelev:2014gla, TheATLAScollaboration} is the technique adopted in the present analyses to estimate the contribution from photonic electrons. 
With a contribution of 80\% to the inclusive electron sample, photonic electrons constitute the main background at $p_\mathrm{T}=0.5$ GeV/$c$ \cite{Acharya:2018upq}. Their contribution decreases with $p_\mathrm{T}$ reaching 25\% at about $3$ GeV/$c$.
The contribution from di-electron decays of light vector mesons ($\rho$, $\omega$ and $\phi$) is negligible compared to the contributions from the photonic sources \cite{Abelev:2012xe}.

Photonic electrons are reconstructed statistically by pairing electron (positron) tracks with opposite charge tracks identified as positrons (electrons), called associated electrons in the following, forming the so-called unlike-sign pairs. The combinatorial background is subtracted using the like-sign invariant mass distribution in the same interval.
Associated electrons are selected with the criteria listed in Table~\ref{tab::associated_track_cuts}, which are intentionally looser than the ones applied for the inclusive electron selection, shown in Table~\ref{tab::track_cuts}, in order to maximise the probability to find the photonic partners. 

\begin{table}[b!]
	\centering
	\caption{Selection criteria for tagging photonic electrons.}
	\begin{tabular}{cccc}
		\toprule
		Associated electron	&	pp	& Pb--Pb  & Pb--Pb \\
			&		&  ($p_\mathrm{T}$ $<$ 3 GeV/$c$) &  ($p_\mathrm{T}$ $>$ 3 GeV/$c$) \\
		\midrule
		$p_\mathrm{T}^{\text{min}}$ (GeV/$c$) & 0.1 & 0.1 & 0.2 \\
		$|y|$ & $<$ 0.8 & $<$ 0.8 & $<$ 0.9 \\
		Number of clusters in TPC & $\geq 60$ & $\geq 60$ & $\geq70$ \\
		TPC clusters  in d$E$/d$x$ calculation & $\geq 60$ & $\geq 60$ & -- \\
		Number of clusters in ITS & $\geq2$ & $\geq2$ & $\geq2$ \\
		
		$|$DCA$_\text{xy}|$ & $<1$ cm & $<1$ cm & $<2.4$ cm \\
		$|$DCA$_\text{z}|$ & $<2$ cm & $<2$ cm & $<3.2$ cm \\
		Found / findable clusters in TPC & $>0.6$ & $>0.6$ & -- \\
		$\chi^2$/d.o.f TPC & $<$ 4 & $<$ 4 & $<$ 4 \\
		$n_{\sigma\textrm{,e}}^{\text{TPC}}$ & $[-3, 3]$ & $[-3, 3]$ & $[-3, 3]$ \\
		\midrule
		$m_{\text{e}^+\text{e}^-}$ (MeV/$c^2$)& $<140$ & $<140$ & $<100$ \\ 
		\bottomrule		
	\end{tabular}
	\label{tab::associated_track_cuts}
\end{table}

Due to the limited acceptance of the detector and the rejection of some associated electrons by applying the mentioned criteria, a certain fraction of photonic pairs is not reconstructed. Therefore, the raw yield of tagged photonic electrons is corrected for efficiency to find the associated electron (positron), the so called tagging efficiency ($\varepsilon_{\text{tag}}$). This is evaluated using Monte Carlo (MC) simulations; pp and Pb--Pb collisions are simulated by the PYTHIA 6 \cite{Sjostrand:2006za} and HIJING \cite{Hijing:ref} event generators, respectively. 
Primary particle generation is followed by particle transport with GEANT3 \cite{Brun:1082634} and a detailed detector response simulation and reconstruction.
The tagging efficiency is defined as the ratio of the number of true reconstructed unlike-sign pair electrons and the number of those generated in the simulations. The simulated $p_\mathrm{T}$ distributions of $\pi^0$ or $\eta$ mesons are weighted in MC to match the measured spectra. In both pp and Pb--Pb collisions, the weighting factor for $\pi^0$ is provided by using the measured distributions of charged pions \cite{collaboration2019production}. 
The weighting factor for $\eta$ mesons is computed using an $m_\text{T}$--scaling approach \cite{Khandai:2011cf,Altenkamper:2017qot}. 
The total tagging efficiency has a monotonic trend. In pp collisions, it starts at $ 0.4$ for $p_\mathrm{T}=0.5$ GeV/$c$ and rises until $p_\mathrm{T}=3$ GeV/$c$, where it flattens at $ 0.7$. In Pb--Pb collisions, it follows the same trend, increasing from $ 0.3$ to $ 0.7$ in the same $p_\mathrm{T}$ range.

It was observed in the previous analysis \cite{Acharya:2018upq} that the contribution from J/$\psi$ decays reaches a maximum of around 5\% in the region $2<p_\mathrm{T}<3$ GeV/$c$ in central Pb--Pb collisions, decreasing to a few percent in more peripheral events. At lower and higher momenta, this contribution quickly decreases and becomes negligible, hence it is not subtracted in the present analyses. The associated systematic uncertainty is taken from similar works \cite{Acharya:2018upq,Adam:2016khe}.
Due to the requirement of hits in both pixel
layers, it was also observed from similar studies in previous measurements \cite{Acharya:2018upq} that the relative contribution from $K_{e{3}}$ decays to the electron background is negligible, hence this contribution is not subtracted in the present analyses. 
Additional sources of background, such as electrons from $W$ and $Z/\gamma^{*}$ decays, are subtracted from the fully corrected and normalised electron yield in Pb--Pb collisions at high $p_\mathrm{T}$. These contributions are obtained from calculations using the POWHEG event generator \cite{Oleari:2010nx} for pp collisions 
and scaling it by $\langle N_\text{coll}\rangle$, assuming $R_\mathrm{AA}=1$.
The contribution from $W$ decays increases
from 1\% at $p_\mathrm{T}=10$ GeV/$c$ to about 20\% at $p_\mathrm{T}=25$ GeV/$c$ in the 0–10\% centrality class, while the $Z$ contribution reaches about $10\%$ at the same transverse momentum.

\subsection{Efficiency correction and normalisation}
After the statistical subtraction of the hadron contamination and the background from photonic electrons, the raw yield of electrons and positrons in bins of $p_\mathrm{T}$ is divided by the number of analysed events ($N_\text{ev}^\text{MB}$), by the 
transverse momentum value at the bin centre $p_\mathrm{T}^\text{centre}$ 
and the bin width $\Delta p_\mathrm{T}$, by the width $\Delta y$ of the covered rapidity interval, by the geometrical acceptance ($\varepsilon^\text{geo}$) times the reconstruction ($\varepsilon^\text{reco}$) and PID efficiencies ($\varepsilon^{\text{eID}}$), and by a factor of two to obtain the charge averaged invariant differential yield, since in the analyses the distinction between positive and negative charges is not done:
\begin{equation}
	\frac{1}{2\pi p_\mathrm{T}}\frac{\text{d}^2N^{e^\pm}}{\text{d}p_\mathrm{T}\text{d}y}=\frac12\frac{1}{2\pi p_\mathrm{T}^\text{centre}}\frac{1}{N_\text{ev}^\text{MB}}\frac{1}{\Delta y\Delta p_\mathrm{T}}\frac{N^{e^\pm}_\text{raw}(p_\mathrm{T})}{(\varepsilon^\text{geo}\times \varepsilon^\text{reco}\times\varepsilon^\text{eID})}\hspace{0.1cm}.
	\label{eq::corrected_yield}
\end{equation}
The production cross section in pp collisions is calculated by multiplying the invariant yield of Eq. \ref{eq::corrected_yield} by the minimum bias trigger cross section at $\sqrt{s}=5.02$ TeV, that is 50.9 $\pm$ 0.9 mb \cite{ALICE-PUBLIC-2018-014}. The per-event yield of electrons from the EMCal triggered sample was scaled to the minimum bias yield by normalisation factors determined with a data-driven method based on the ratio of the energy distributions of EMCal clusters for the two triggers, as described in Ref. \cite{Adam:2016khe}. The normalisation is 64.5 $\pm$ 0.5 in 0--10\% and 246 $\pm$ 2.6 in 30--50\% centrality intervals, respectively.

The efficiencies are determined using specific MC simulations, where every collision event is produced with at least either a $c\overline{c}$ or $b\overline{b}$ pair and heavy-flavour hadrons are forced to decay semileptonically to electrons \cite{Acharya:2018upq, Adam:2016khe}. 
The  underlying  Pb–Pb  events  were  simulated  using  the  HIJING 
generator \cite{Hijing:ref} and heavy-flavour signals were added using the PYTHIA 6 
generator \cite{Sjostrand:2006za}. 
The efficiency of reconstructing electrons from semileptonic heavy-flavour hadron decays 
is about $ 20\%$ at $p_\mathrm{T}=0.5$ GeV/$c$, then it increases with $p_\mathrm{T}$ up to $ 58\%$ in pp collisions. 
In Pb--Pb collisions, it follows the same trend, increasing from $5\%$ to $10\%$ in the same $p_\mathrm{T}$ range.

\subsection{Systematic uncertainties}

The overall systematic uncertainties on the $p_\mathrm{T}$ spectra are calculated summing in quadrature the different contributions, which are assumed to be uncorrelated. They are summarised in Table \ref{tab:systematic} and discussed in the following.

The systematic uncertainties on the total reconstruction efficiency arising from the comparison between MC and data are estimated by varying
the track selection and PID requirements around the default values chosen in the analyses. 
The analysis is repeated with tighter and looser conditions with respect to the default selection criteria
and the systematic uncertainty is calculated as the root mean square (RMS) of the distribution of the resulting corrected yields (or cross sections in pp) in each centrality and $p_\mathrm{T}$ interval. 
The systematic uncertainty estimated in pp collisions is less than 2\%, while in Pb--Pb collisions it reaches a maximum value of 4\% in 0--10\% centrality class for $p_\mathrm{T}<0.9$ GeV/$c$. 

Similarly, the systematic uncertainty arising from the photonic-electron subtraction technique is estimated as the RMS of the distribution of yields obtained by varying the selection criteria listed in Table \ref{tab::associated_track_cuts}.
In pp collisions this contribution has a maximum of 4\% for $0.5<p_\mathrm{T}<0.7$ GeV/$c$ and then it gradually decreases with increasing $p_\mathrm{T}$, while in the 0--10\% Pb--Pb centrality class it is the dominant source of systematic uncertainty, being 13\% in the first $p_\mathrm{T}$ interval. This systematic uncertainty mainly arises when the invariant mass criterion on the photonic pairs is varied and it reflects the large contribution of photonic electrons in the low-$p_\mathrm{T}$ region.  

In order to further test the robustness of the photonic electron tagging,
the requirement on the number of clusters for electron candidates in the SPD is 
relaxed in order to increase
the fraction of electrons coming from photon conversions in the detector material. A variation of 3\% is observed for the measured pp cross section in the full $p_\mathrm{T}$ range, while in central Pb--Pb collisions the observed deviation amounts to 10\% for $0.5<p_\mathrm{T}<0.7$ GeV/$c$, decreasing with increasing $p_\mathrm{T}$.
This systematic uncertainty is less relevant in semi-central collisions, and it is compatible with the variation determined in pp measurements for $1.5<p_\mathrm{T}<3$ GeV/$c$.

In addition, the systematic uncertainty related to the subtraction of the background electrons from W and $Z/\gamma^{*}$ is estimated by propagating 15\% of uncertainty, 
which covers the possible difference between the measurements and the theoretical calculations
\cite{Aad:2014qxa, Aad:2014qja, Sirunyan:2018owv}.
The uncertainty from the subtraction on the final result, which is relevant only at high $p_\mathrm{T}$, is less than 4\% for electrons from semileptonic heavy-flavour hadron decays in central (0--10\%) Pb--Pb collisions for 24$<p_{\textrm{T}}<$ 26 GeV/$c$, and less than 1\% in other centrality classes for the same $p_\mathrm{T}$ interval.
In the pp analysis, a 5\% systematic uncertainty is found while varying the selection criterion in the TPC for $p_\mathrm{T}>8$ GeV/$c$ due to the increasing relative amount of hadrons. An additional systematic uncertainty of 5\%, related to the precision of the estimated hadron contamination, is assigned for $p_\mathrm{T}>8$ GeV/$c$.
In Pb--Pb collisions, a 10\% systematic uncertainty is assigned for $p_\mathrm{T}>12$ GeV/$c$ due to the variation of electron identification in the TPC, while this contribution is within 5\% at lower $p_\mathrm{T}$. Moreover, a 6\% uncertainty is assigned due to the $E/p$ selection criterion. Finally, for $p_\mathrm{T}<3$ GeV/$c$, different functional forms are used for the parametrisation of the pion contribution in the fitting procedure adopted to evaluate the hadron contamination. A systematic uncertainty of about 6\% is assigned for $p_\mathrm{T}<3$ GeV/$c$ in the 0--10\% centrality class, while this contribution decreases for more peripheral collisions.

In the pp (Pb--Pb) analysis, a systematic uncertainty of about 2\% (3\%) is assigned due to the incomplete knowledge of the efficiency in matching tracks reconstructed in the ITS and TPC and another 2\% (5\%) for the track matching between TPC and TOF. 

The effects due to the presence of 
non-uniformity in the correction for the space-charge distortion in the TPC drift volume
or irregularities in the detector coverage are then evaluated by repeating the analysis in different geometrical regions. In pp collisions, a maximum systematic uncertainty of 5\% is estimated from varying the pseudorapidity range used for the cross section measurement. The same value is assigned in the 30--50\% and 60--80\% Pb--Pb centrality intervals, while a 10\% systematic uncertainty is assigned for $0.5<p_\mathrm{T}<0.7$~GeV/$c$ in the 0--10\% centrality interval, due to the larger sensitivity to the electrons from photon conversions. An additional uncertainty of 10\% for $p_\mathrm{T}<1$ GeV/$c$ and of 5\% up to $p_\mathrm{T}=3$ GeV/$c$ is estimated from varying the azimuthal region in central Pb--Pb collisions. Furthermore, the analysis of Pb--Pb collisions is repeated using different interaction rate regimes. A 5\% deviation is observed at low $p_\mathrm{T}$ in central Pb--Pb collisions when selecting only high ($>$ 5 kHz) or low ($<$ 5 kHz) interaction rate events.

The uncertainty from the EMCal trigger normalisation in Pb--Pb collisions at $p_\mathrm{T}>12$ GeV/$c$ is estimated as the RMS of the rejection factor values computed at different transverse momenta \cite{Adam:2016khe}. The RMS is 
4\%
and assigned as the systematic uncertainty.

The uncertainties on the $R_\mathrm{AA}$ normalisation are the quadratic sum of the uncertainties on the average nuclear overlap functions in Table \ref{tab::num_evs}, the normalisation uncertainty due to the luminosity and the uncertainty related to the determination of the centrality intervals, which reflects the uncertainty on the fraction of the hadronic cross section used in the Glauber fit to determine the centrality \cite{Acharya:2018hre,Adam:2015sza}.

\begin{table}[ht!]
    \centering
    \caption{ Contributions to the systematic uncertainties on the cross section (yield) of electrons from heavy-flavour
  hadron decays in pp (Pb--Pb) collisions, quoted for the transverse momentum intervals $0.5<p_\mathrm{T}<0.7$ GeV/$c$ and $8<p_\mathrm{T}<10$ GeV/$c$.
  These $p_\mathrm{T}$ intervals are listed because the detectors used for particle identification in the two cases are different. In addition, they also represent the first and the last $p_\mathrm{T}$ intervals in common for the centrality classes in Pb--Pb collisions, as well as for the pp cross section. At higher $p_\mathrm{T}$ the uncertainties are generally lower, apart from the one related to the electroweak background, which stays below 4\%.
  The uncertainties quoted with * are not summed in quadrature together with the others, because they are the $R_\mathrm{AA}$ normalization uncertainties.
  }
    \begin{tabular}{l|cc|cc|cc|cc}
         & 
         \multicolumn{2}{|c|}{pp} & 
         \multicolumn{2}{c|}{Pb--Pb (0--10\%)} &
         \multicolumn{2}{c|}{Pb--Pb (30--50\%)} & \multicolumn{2}{c}{Pb--Pb (60--80\%)}  \\
    \toprule
    $\boldsymbol{p_\mathrm{T}}$ {(GeV/$c$)} & 
        0.5--0.7 & 8--10 &    
        0.5--0.7 & 8--10 &
        0.5--0.7 & 8--10 &
        0.5--0.7 & 8--10 \\
    \midrule
    Track selections &
        1\% & 1\% &
        4\% & 2\% &
        1\% & 2\% &
        2\% & 2\% \\
    Photonic
    tagging &
        4\% & -- &
        13\%&  4\% &
        7\% & 4\% &
        7\% & 4\% \\
    SPD hit requirement &
        3\% & 3\% &
        10\% & -- &
        -- & -- & 
        -- & -- \\
    J/$\psi$$\rightarrow$e 
        &
        -- & -- & 
        2\% & -- &
        2\% & -- & 
        2\% & -- \\
    W$\rightarrow$e 
    &
        -- & -- & 
        -- & $<$4\% &
        -- & $<$1\% & 
        -- & $<$1\% \\
    Z/$\gamma\rightarrow$e 
    &
        -- & -- & 
        -- & $<$1\% &
        -- & $<$1\% & 
        -- & $<$1\% \\
    $n_{\sigma\textrm{,e}}^{\textrm{TPC}}$ selection &
        -- & 5\% &
        -- & 5\% &
        -- & 5\% &
        -- & 2\% \\
    $E/p$ selection &
        -- & -- &
        -- & 6\% &
        -- & 6\% &
        -- & 6\% \\
    Hadron contamination &
        -- & 5\% &
        6\% & -- &
        2\% & -- &
        -- & -- \\
    ITS--TPC matching &
        2\% & 2\% &
        2\% & 2\% &
        2\% & 2\% &
        2\% & 2\% \\
    TPC--TOF matching &
        2\% & -- &
        3\% & -- &
        -- & -- &
        -- & -- \\
    $\eta$ &
        5\% & 4\% &
        10\% & -- &
        5\% & -- &
        5\% & -- \\
    $\varphi$ & 
        -- & -- &
        10\% & -- &
        -- & -- &
        -- & -- \\
    Interaction rate &
        -- & -- & 
        5\% & -- &
        -- & -- & 
        -- & -- \\
    \midrule
    Centrality limit* & \multicolumn{2}{|c}{--} & \multicolumn{2}{|c}{$<1$\%} & 
    \multicolumn{2}{|c}{$2$\%} & \multicolumn{2}{|c}{$3$\%} \\
    \midrule
    Luminosity* & \multicolumn{2}{c}{2.1\%} & \multicolumn{2}{|c}{--} & \multicolumn{2}{|c}{--} & \multicolumn{2}{|c}{--} \\
    \midrule
    Total uncertainty &
        9\% & 9\% &
        24\% & 9\% &
        9\% & 9\% &
        9\% & 8\%
        \\
    \bottomrule
    \end{tabular}
    \label{tab:systematic}
\end{table}
\section{Results}
\subsection{$p_\mathrm{T}$-differential cross section in pp collisions and invariant yield in Pb--Pb collisions}

\begin{figure}[ht!]
    \centering
    %
    %
    \includegraphics[scale=0.6]{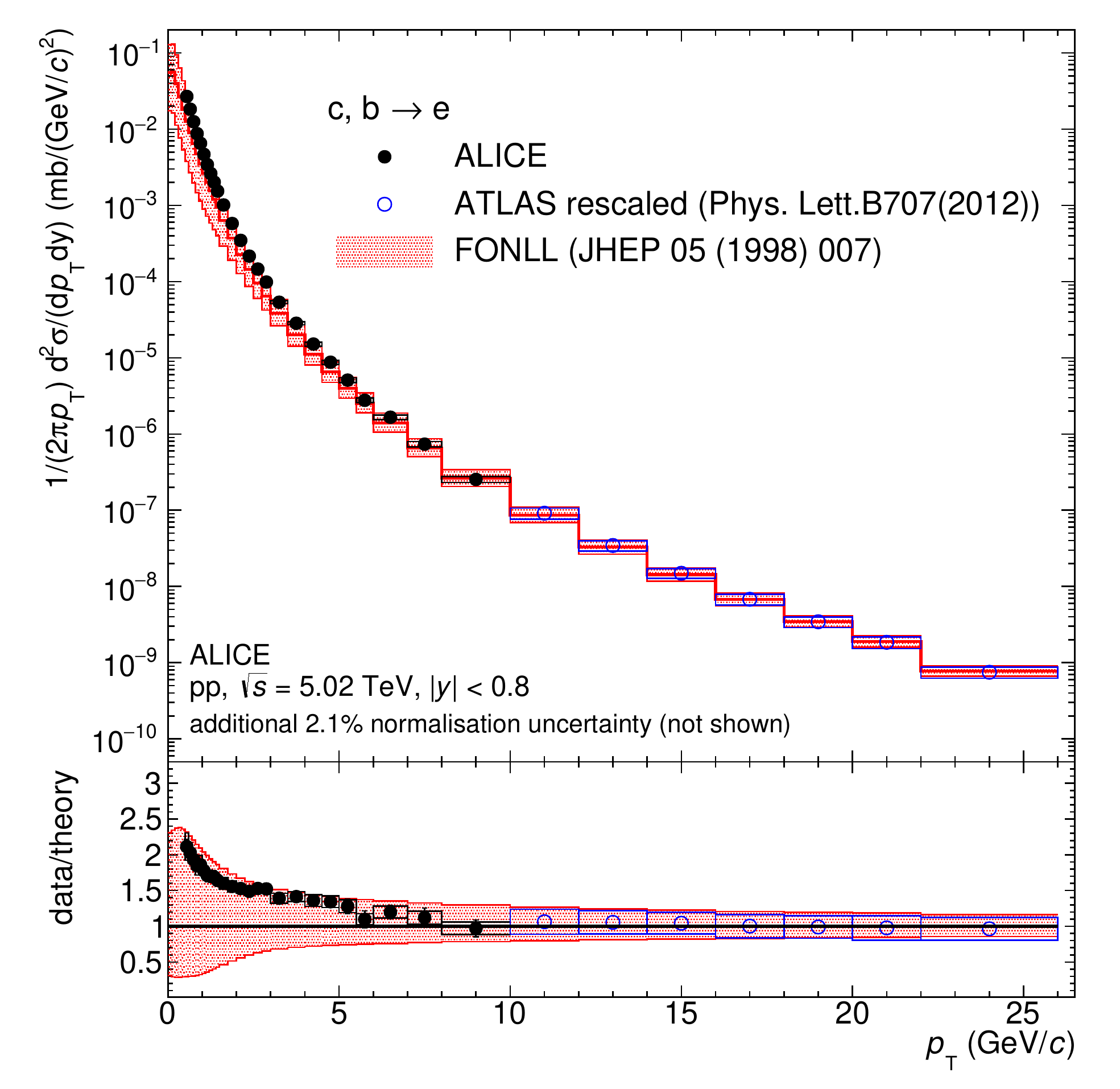}
    \caption{$p_\mathrm{T}$-differential invariant production cross section of electrons from semileptonic heavy-flavour hadron decays 
    in pp collisions at $\sqrt{s}=5.02$~TeV. The measurement is compared with the FONLL calculation \cite{Cacciari:1998it}. In the bottom panel, the ratios with respect to the central values of the FONLL calculation are shown. An additional 2.1\% normalisation uncertainty, due to the measurement of the minimum bias triggered cross section \cite{ALICE-PUBLIC-2018-011}, is not shown in the results.}
    \label{fig:results_pp_cross_section}
\end{figure}

The $p_\mathrm{T}$-differential production cross section of electrons from semileptonic heavy-flavour hadron decays 
in pp collisions at $\sqrt{s}=5.02$ TeV is shown in Fig. \ref{fig:results_pp_cross_section}.
The data in the region $0.5<p_\mathrm{T}<10$ GeV/$c$ is compared with the Fixed-Order-Next-to-Leading-Log (FONLL) \cite{Cacciari:1998it} pQCD calculation. The uncertainties of the FONLL calculations
(dashed area) reflect different choices for the charm and beauty quark masses, the factorisation and renormalisation scales as well as the uncertainty on the set of parton distribution functions (PDF) used in the pQCD calculation (CTEQ6.6 \cite{Nadolsky:2008zw}). The measured cross section is close to the upper edge of the theoretical prediction up to $p_\mathrm{T}\simeq 5$ GeV/$c$, as observed in pp collisions at $\sqrt{s}=2.76$ and 7 TeV \cite{Acharya:2018upq, Abelev:2014gla, Abelev:2012xe
}, while at higher $p_\textrm{T}$, where electrons from semileptonic beauty hadron decays are expected to dominate, the measurement is close to the mean value of the FONLL prediction. 

\begin{figure}[ht!]
    \centering
    %
    %
    \includegraphics[width=0.75\textwidth]{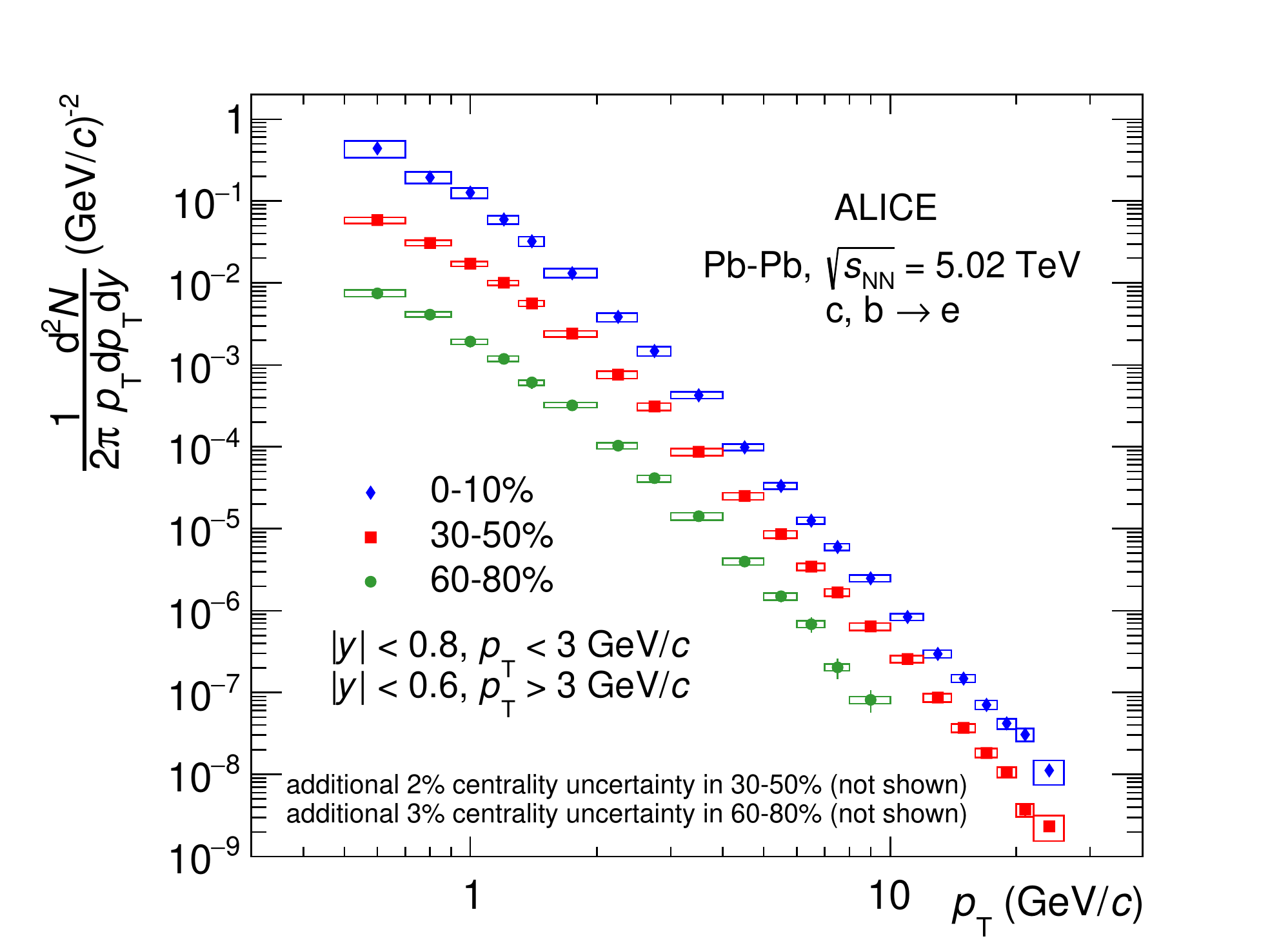}
    \caption{$p_\mathrm{T}$-differential invariant yield   
    in central (0--10\%), semi-central (30--50\%), and peripheral (60--80\%) Pb--Pb collisions at $\sqrt{s_\mathrm{NN}}=5.02$ TeV.}
    \label{fig:results_PbPb_yields}
\end{figure}

The $p_\mathrm{T}$-differential invariant yield of electrons from semileptonic heavy-flavour hadron decays  
measured in central (0--10\%), semi-central (30--50\%), and peripheral (60--80\%) Pb--Pb collisions at $\sqrt{s_\mathrm{NN}}=5.02$~TeV is shown in Fig. \ref{fig:results_PbPb_yields}. The measurements are performed in the $p_\mathrm{T}$ interval 0.5--26 GeV/$c$ in the 0--10\% and in the 30--50\% centrality intervals, and only up to $p_\mathrm{T}$ = 10~GeV/$c$ in the 60--80\% centrality class due to limited statistics in Pb-Pb data recorded in 2015. 


\subsection{Nuclear modification factor}

\begin{figure}[]
    \centering
    %
    %
    %
    %
    \includegraphics[width=1.0\textwidth]{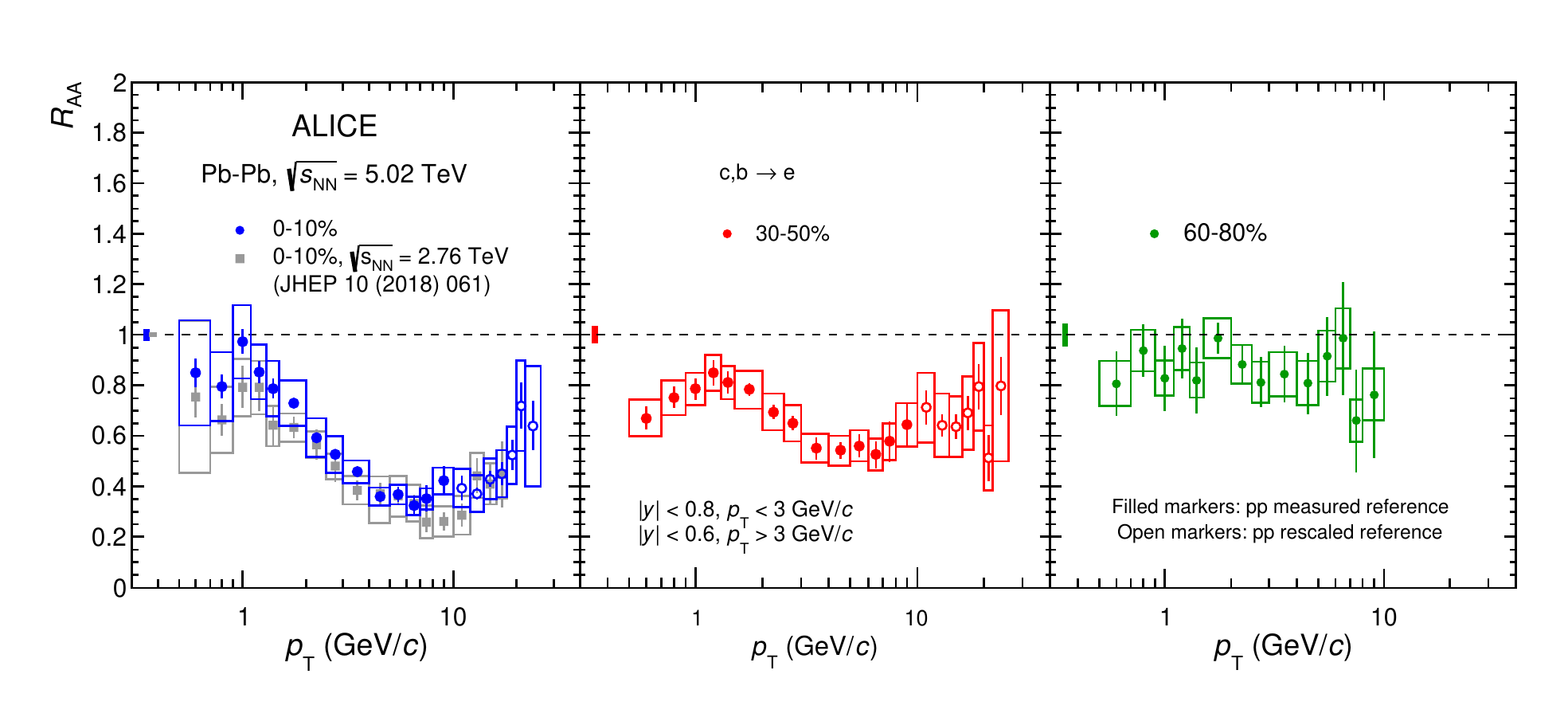}
    \caption{Nuclear modification factor of electrons from semileptonic heavy-flavour hadron decays  
    measured in the three centrality intervals in Pb--Pb collisions at $\sqrt{s_\mathrm{NN}}=5.02$ TeV.}
    \label{fig:results_RAA_diff_centr}
\end{figure}

The nuclear modification factor of electrons from semileptonic heavy-flavour hadron decays 
measured in central (0--10\%), semi-central (30--50\%), and peripheral (60--80\%) Pb--Pb collisions at $\sqrt{s_\mathrm{NN}}=5.02$ TeV is shown in Fig. \ref{fig:results_RAA_diff_centr}. The measured cross section in pp collisions at $\sqrt{s}=5.02$ TeV (Fig. \ref{fig:results_pp_cross_section}) is used as a reference up to $p_\mathrm{T}=10$ GeV/$c$. For $p_{\textrm{T}}>10$ GeV/$c$, the reference is obtained by a $p_\mathrm{T}$-dependent scaling of the measurement at $\sqrt{s}=7$ TeV by the ATLAS collaboration \cite{Aad:2011rr} with the ratio of the cross section at the two collision energies computed with the FONLL calculation \cite{Averbeck:2011ga}. 
This ratio is performed by considering the different rapidity coverage of the ATLAS measurement ($|y|<2$ excluding $1.37<|y|<1.52$).
The systematic uncertainties of the cross section at $\sqrt{s}=5.02$ TeV range from 13\% to 18\% depending on the $p_\mathrm{T}$ interval, and they are computed as the propagation of the uncertainties associated with FONLL calculations at $\sqrt{s}=5.02$ TeV and $\sqrt{s}=7$ TeV and the systematic uncertainties of the ATLAS measurement. The statistical uncertainties are from the ATLAS measurement.

Statistical and systematic uncertainties of the $p_\mathrm{T}$-differential yields and cross sections in Pb–Pb and pp collisions, respectively, are propagated as uncorrelated uncertainties. 
The uncertainties on the $R_\mathrm{AA}$ normalisation are reported in Fig. \ref{fig:results_RAA_diff_centr} as boxes at unity.
The measured $R_\mathrm{AA}$ shows a clear dependence on the collision centrality, since in most central events it reaches a minimum of about $0.3$ around $p_\mathrm{T}=7$~GeV/$c$, while moving to more peripheral Pb--Pb collisions the $R_\mathrm{AA}$ gets closer to unity at $p_\mathrm{T}>3$ GeV/$c$. Such a suppression is not observed in proton-lead collisions at the same energy where the QGP is not expected to be formed and the nuclear modification factor is consistent with unity \cite{Adam:2016wyz, Adam:2015qda, Abelev:2014hha}. Thus the suppression of electron production is due to final-state effects, 
such as partonic energy loss in the medium. Since electrons from semileptonic beauty decays are expected to dominate the spectrum 
at high $p_{\textrm{T}}$ while charm production dominates at low $p_{\textrm{T}}$ \cite{Adam:2016wyz}, the measurements show that charm and beauty quarks lose energy in the medium.
The centrality dependence of the $R_\textrm{AA}$ is compatible with the hypothesis of a partonic energy loss dependence on medium density, being larger in a hotter and denser QGP, like the one created in the most central collisions. In addition, it reflects a path-length dependence of energy loss. Moreover, it has been shown in Refs. \cite{Acharya:2018njl,Morsch:2017brb} that a centrality selection bias is present in peripheral Pb--Pb collisions which reduces the $R_\mathrm{AA}$ below unity even in the absence of any nuclear modification effects. This effect may be responsible for a significant part of the apparent suppression seen in the $R_\mathrm{AA}$ of electrons from semileptonic heavy-flavour hadron decays in the 60-80\% centrality class.

For $p_\mathrm{T}<7$~GeV/$c$, the $R_\mathrm{AA}$ of electrons from semileptonic heavy-flavour hadron decays 
increases with decreasing $p_\mathrm{T}$ as a consequence of the scaling of the total heavy-flavour yield with the number of binary collisions among nucleons in Pb--Pb collisions. 
On the other hand, the nuclear modification factor at low $p_\mathrm{T}$ does not rise above unity.
This kinematic region is sensitive to the effects of nuclear shadowing: the depletion of parton densities in nuclei at low Bjorken $x$ values can reduce the heavy-quark production cross section per binary collision in Pb–Pb with respect to the pp case \cite{Acharya:2018upq}. This initial-state effect is studied in p–Pb collisions, however, the present uncertainties on the $R_\mathrm{pPb}$ measurement do not allow quantitative
conclusions on the modification of the PDF in nuclei in the low $p_\mathrm{T}$ region to be made \cite{Adam:2015qda}. Furthermore, the amount of electrons from semileptonic heavy-flavour hadron decays is reduced due to the presence of hadrochemistry effects. For example, $\Lambda_{\rm c}^+$ baryons decay into electrons with a branching ratio of $5\%$, while for the D mesons the branching ratio is less than $10\%$. Since in Pb--Pb collisions more charm quarks might hadronize into baryons \cite{Acharya:2018ckj}, this effect reduces the total amount of electrons from semileptonic heavy-flavour hadron decays.
Additional effects, such as collective motion induced by the medium, also have an influence on the measured $R_\mathrm{AA}$. Also, it has been observed that the radial flow can provoke an additional yield enhancement at intermediate $p_\mathrm{T}$
\cite{He:2011qa, He:2012df, Gossiaux:2010yx, Gossiaux:2012ya}.
In this case, the radial flow pushes up slow particles to higher momenta, causing a small increase in the nuclear modification factor around $p_\mathrm{T}=1$ GeV/$c$.

It should be noted
that the $R_\mathrm{AA}$ measurements in the most central collisions at $\sqrt{s_\mathrm{NN}}$ = 2.76 TeV \cite{Acharya:2018upq} and 5.02 TeV are compatible within uncertainties, as shown in Fig. \ref{fig:results_RAA_diff_centr}. This effect was predicted by the Djordjevic model \cite{Djordjevic:2015hra}, and it results from the combination of a higher medium temperature at 5.02 TeV, which would decrease the $R_\mathrm{AA}$ by about 10\%, with a harder $p_\mathrm{T}$ distribution of heavy quarks at 5.02 TeV, which would increase the $R_\mathrm{AA}$ by about 5\% if the medium temperature were the same as at 2.76 TeV. An analogous behaviour between the measured $R_\mathrm{AA}$ at the two energies is also observed for the D mesons \cite{Acharya:2018hre}.


\subsection{Comparison with model predictions}
\begin{figure}[ht!]
    \centering
    \subfloat[][0--10\%]
    %
    %
    {\includegraphics[width=0.45\textwidth]{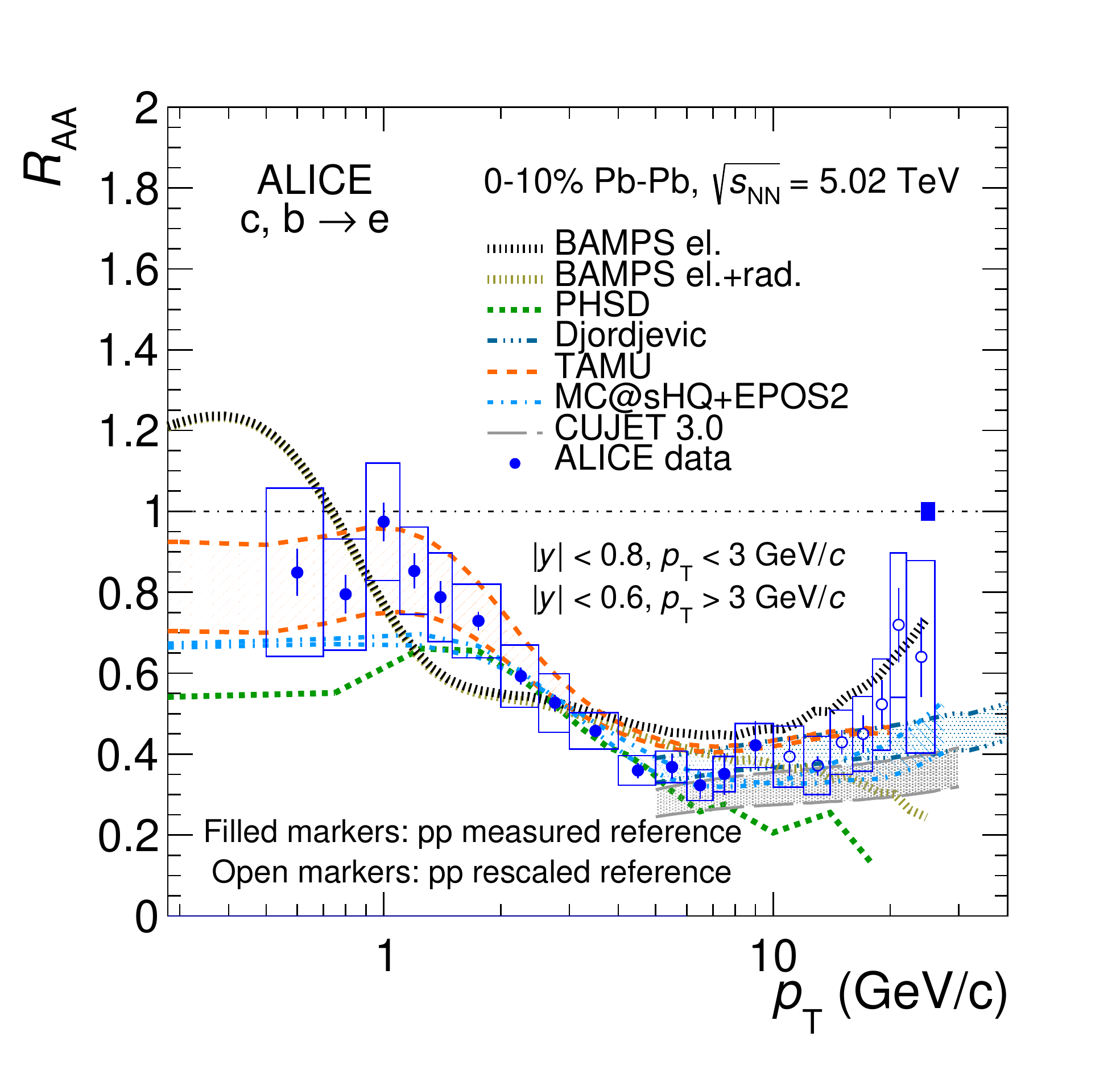}}
    \subfloat[][30--50\%]
    %
    %
    {\includegraphics[width=0.45\textwidth]{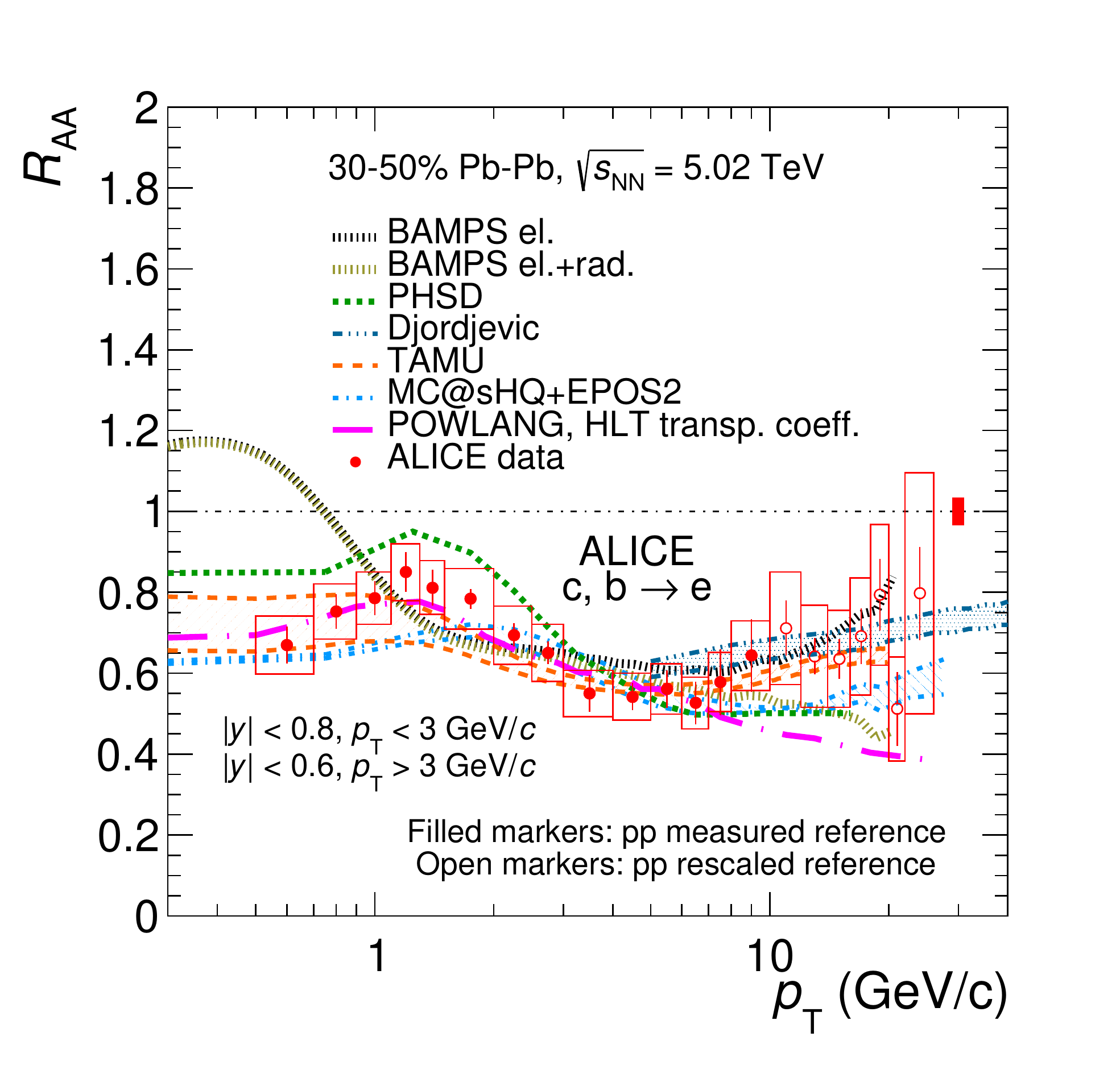}}
    \caption{Nuclear modification factor of electrons semileptonic from heavy-flavour hadron decays measured in 0--10\% and 30--50\% centrality in Pb--Pb collisions at $\sqrt{s_\mathrm{NN}}=5.02$ TeV compared with model predictions \cite{Xu:2015bbz, Uphoff:2014hza, Song:2015ykw, Djordjevic:2015hra, He:2014cla, Djordjevic:2014, Beraudo:2014boa, Nahrgang:2013xaa}.}
    \label{fig:results_RAA_comp_theory}
\end{figure}

In Fig. \ref{fig:results_RAA_comp_theory} the measured $R_\mathrm{AA}$ in the 0--10\% (left panel) and 30--50\% (right panel) centrality intervals are compared with model calculations \cite{Xu:2015bbz, Uphoff:2014hza, Song:2015ykw, Djordjevic:2015hra, He:2014cla, Djordjevic:2014, Beraudo:2014boa, Nahrgang:2013xaa}. The model calculations take into account different hypotheses about mass dependence of energy loss processes, transport dynamics, charm and beauty quark interactions with the QGP constituents, hadronisation mechanisms of heavy quarks in the plasma, and heavy-quark production cross section in nucleus--nucleus collisions. 

Most of the models provide a fair description of the data in the region $p_\mathrm{T}$ $<$ 5 GeV/$c$ in both 
centrality classes,
except for BAMPS \cite{Uphoff:2014hza}.
The predictions from the MC@sHQ+EPOS2 \cite{Nahrgang:2013xaa}, PHSD \cite{Song:2015ykw}, TAMU \cite{He:2014cla}, and POWLANG \cite{Beraudo:2014boa} models
also include nuclear modification of the parton distribution functions, which is necessary to predict the observed suppression of the $R_\mathrm{AA}$ at low $p_\mathrm{T}$.  
The following observations about the comparison with model calculations are fully in agreement with what is observed in the $R_\mathrm{AA}$ measurements of D mesons \cite{Acharya:2018hre}.

The nuclear modification factor for central Pb--Pb collisions is well described by the TAMU \cite{He:2014cla} prediction at  $p_\mathrm{T}<3$ GeV/$c$ within the uncertainties related to the shadowing effect on charm quarks. 
However, this model tends to overestimate the $R_\mathrm{AA}$ for $p_{\mathrm{T}}>3$ GeV/$c$, probably due to the missing implementation of the radiative energy loss in the model,
which may become the dominant energy loss mechanism at high $p_\mathrm{T}$. 

The agreement with TAMU \cite{He:2014cla} at low $p_\mathrm{T}$, on the other hand, confirms the dominance of elastic collisions at low momenta, together with the importance of the inclusion of shadowing effects in the model calculations \cite{Eskola:2009uj},
which reduce the total heavy-flavour production in Pb--Pb collisions with respect to an expectation from the binary scaling.

In semi-central Pb--Pb collisions 
the TAMU \cite{He:2014cla} and POWLANG \cite{Beraudo:2014boa}  predictions are close to the lower edge of the uncertainties of the measured $R_\mathrm{AA}$
for $p_\mathrm{T}<3$ GeV/$c$. The latter calculation describes the data better up to $p_\mathrm{T}\simeq 8$ GeV/$c$, while the former provides a good description even at higher transverse momenta. The CUJET3.0 \cite{Xu:2015bbz} and Djordjevic \cite{Djordjevic:2015hra, Djordjevic:2014} models provide a good description of the $R_\mathrm{AA}$ within the uncertainties in both centrality intervals for $p_\mathrm{T}$ $>$ 5 GeV/$c$, suggesting that the dependence of radiative energy loss on the path length in the hot and dense medium is well understood. 

\section{Conclusions}
The invariant yield of electrons from semileptonic heavy-flavour hadron decays was measured in central (0--10\%), semi-central (30--50\%), and peripheral (60--80\%) Pb--Pb collisions at $\sqrt{s_\mathrm{NN}}=5.02$ TeV. The measurement of the nuclear modification factor in all the centrality classes for $p_\mathrm{T}<10$ GeV/$c$ is provided using as reference the cross section measured in pp collisions at the same centre-of-mass energy. The systematic uncertainties of this measurement are reduced by a factor of about 2 compared to the published reference in pp collisions at $\sqrt{s}=2.76$ TeV \cite{Acharya:2018upq} and the measured cross section is close to the upper edge of the FONLL uncertainty band. At higher $p_\mathrm{T}$ the reference is obtained by a $p_\mathrm{T}$-dependent scaling of the measurement at $\sqrt{s}=7$ TeV by the ATLAS collaboration \cite{Aad:2011rr} with the ratio of the cross section at the two collision energies computed with the FONLL calculation \cite{Averbeck:2011ga}.
As in the Pb--Pb analysis at $\sqrt{s_\mathrm{NN}}=2.76$ TeV \cite{Acharya:2018upq, Adam:2016khe}, the main source of background electrons, constituted by photonic electrons, is removed via the photonic tagging method. In addition, compared with the measurements performed in pp and Pb--Pb collisions at $2.76$ TeV, the $p_\mathrm{T}$ range is extended, 
and an additional centrality class is added.

The measured $R_\mathrm{AA}$ confirms the evidence of a strong suppression with respect to what is expected from a simple binary scaling for large $p_\mathrm{T}$. This is a clear signature of the medium induced energy loss on heavy quarks traversing the QGP produced in heavy-ion collisions.

The measurement of electrons from semileptonic heavy-flavour hadron decays in different centrality classes exhibits the dependence of energy loss on the path length and energy density in the hot and dense medium.
The $R_\mathrm{AA}$ at high $p_\mathrm{T}$ (above 5 GeV/$c$) is fairly described in the 0--10\% and 30--50\% centrality intervals by model calculations that include both radiative and collisional energy loss.
This indicates that the centrality dependence of radiative energy loss is theoretically understood.
Further investigations and measurement of electrons from semileptonic decays of beauty hadrons will give more information about the mass dependence of the energy loss in the heavy-flavour sector.

With the good precision of the results presented here, the Pb--Pb data exhibit their sensitivity to the modification of the PDF in nuclei, like nuclear shadowing, which causes a suppression of the heavy-quark production
in heavy-ion collisions. 
The implementation of the nuclear modification of the PDF in theoretical calculations is a necessary ingredient in order for the model predictions to correctly describe the measured $R_\mathrm{AA}$
\cite{Acharya:2018upq}. %
%

\newenvironment{acknowledgement}{\relax}{\relax}
\begin{acknowledgement}
\section*{Acknowledgements}

The ALICE Collaboration would like to thank all its engineers and technicians for their invaluable contributions to the construction of the experiment and the CERN accelerator teams for the outstanding performance of the LHC complex.
The ALICE Collaboration gratefully acknowledges the resources and support provided by all Grid centres and the Worldwide LHC Computing Grid (WLCG) collaboration.
The ALICE Collaboration acknowledges the following funding agencies for their support in building and running the ALICE detector:
A. I. Alikhanyan National Science Laboratory (Yerevan Physics Institute) Foundation (ANSL), State Committee of Science and World Federation of Scientists (WFS), Armenia;
Austrian Academy of Sciences, Austrian Science Fund (FWF): [M 2467-N36] and Nationalstiftung f\"{u}r Forschung, Technologie und Entwicklung, Austria;
Ministry of Communications and High Technologies, National Nuclear Research Center, Azerbaijan;
Conselho Nacional de Desenvolvimento Cient\'{\i}fico e Tecnol\'{o}gico (CNPq), Financiadora de Estudos e Projetos (Finep), Funda\c{c}\~{a}o de Amparo \`{a} Pesquisa do Estado de S\~{a}o Paulo (FAPESP) and Universidade Federal do Rio Grande do Sul (UFRGS), Brazil;
Ministry of Education of China (MOEC) , Ministry of Science \& Technology of China (MSTC) and National Natural Science Foundation of China (NSFC), China;
Ministry of Science and Education and Croatian Science Foundation, Croatia;
Centro de Aplicaciones Tecnol\'{o}gicas y Desarrollo Nuclear (CEADEN), Cubaenerg\'{\i}a, Cuba;
Ministry of Education, Youth and Sports of the Czech Republic, Czech Republic;
The Danish Council for Independent Research | Natural Sciences, the VILLUM FONDEN and Danish National Research Foundation (DNRF), Denmark;
Helsinki Institute of Physics (HIP), Finland;
Commissariat \`{a} l'Energie Atomique (CEA), Institut National de Physique Nucl\'{e}aire et de Physique des Particules (IN2P3) and Centre National de la Recherche Scientifique (CNRS) and R\'{e}gion des  Pays de la Loire, France;
Bundesministerium f\"{u}r Bildung und Forschung (BMBF) and GSI Helmholtzzentrum f\"{u}r Schwerionenforschung GmbH, Germany;
General Secretariat for Research and Technology, Ministry of Education, Research and Religions, Greece;
National Research, Development and Innovation Office, Hungary;
Department of Atomic Energy Government of India (DAE), Department of Science and Technology, Government of India (DST), University Grants Commission, Government of India (UGC) and Council of Scientific and Industrial Research (CSIR), India;
Indonesian Institute of Science, Indonesia;
Centro Fermi - Museo Storico della Fisica e Centro Studi e Ricerche Enrico Fermi and Istituto Nazionale di Fisica Nucleare (INFN), Italy;
Institute for Innovative Science and Technology , Nagasaki Institute of Applied Science (IIST), Japanese Ministry of Education, Culture, Sports, Science and Technology (MEXT) and Japan Society for the Promotion of Science (JSPS) KAKENHI, Japan;
Consejo Nacional de Ciencia (CONACYT) y Tecnolog\'{i}a, through Fondo de Cooperaci\'{o}n Internacional en Ciencia y Tecnolog\'{i}a (FONCICYT) and Direcci\'{o}n General de Asuntos del Personal Academico (DGAPA), Mexico;
Nederlandse Organisatie voor Wetenschappelijk Onderzoek (NWO), Netherlands;
The Research Council of Norway, Norway;
Commission on Science and Technology for Sustainable Development in the South (COMSATS), Pakistan;
Pontificia Universidad Cat\'{o}lica del Per\'{u}, Peru;
Ministry of Science and Higher Education and National Science Centre, Poland;
Korea Institute of Science and Technology Information and National Research Foundation of Korea (NRF), Republic of Korea;
Ministry of Education and Scientific Research, Institute of Atomic Physics and Ministry of Research and Innovation and Institute of Atomic Physics, Romania;
Joint Institute for Nuclear Research (JINR), Ministry of Education and Science of the Russian Federation, National Research Centre Kurchatov Institute, Russian Science Foundation and Russian Foundation for Basic Research, Russia;
Ministry of Education, Science, Research and Sport of the Slovak Republic, Slovakia;
National Research Foundation of South Africa, South Africa;
Swedish Research Council (VR) and Knut \& Alice Wallenberg Foundation (KAW), Sweden;
European Organization for Nuclear Research, Switzerland;
Suranaree University of Technology (SUT), National Science and Technology Development Agency (NSDTA) and Office of the Higher Education Commission under NRU project of Thailand, Thailand;
Turkish Atomic Energy Agency (TAEK), Turkey;
National Academy of  Sciences of Ukraine, Ukraine;
Science and Technology Facilities Council (STFC), United Kingdom;
National Science Foundation of the United States of America (NSF) and United States Department of Energy, Office of Nuclear Physics (DOE NP), United States of America.    
\end{acknowledgement}

\bibliographystyle{utphys}
\bibliography{biblio}

\providecommand{\href}[2]{#2}\begingroup\raggedright\begin{thebibliography}{10}

\bibitem{bib::Alice_performance}
{\bfseries ALICE} Collaboration, B.~Abelev {\em et~al.}, ``{Performance of the
  ALICE Experiment at the CERN LHC},''
  \href{http://dx.doi.org/10.1142/S0217751X14300440}{{\em Int.J.Mod.Phys.}
  {\bfseries A29} (2014) 1430044},
\href{http://arxiv.org/abs/1402.4476}{{\ttfamily arXiv:1402.4476 [nucl-ex]}}.

\bibitem{bib::intro_thermalization}
F.-M. Liu and S.-X. Liu, ``{Quark-gluon plasma formation time and direct
  photons from heavy ion collisions},''
  \href{http://dx.doi.org/10.1103/PhysRevC.89.034906}{{\em Phys. Rev.}
  {\bfseries C89} no.~3, (2014) 034906},
\href{http://arxiv.org/abs/1212.6587}{{\ttfamily arXiv:1212.6587 [nucl-th]}}.

\bibitem{bib::Averbeck}
R.~Averbeck, ``{Heavy-flavor production in heavy-ion collisions and
  implications for the properties of hot QCD matter},''
  \href{http://dx.doi.org/10.1016/j.ppnp.2013.01.001}{{\em Prog. Part. Nucl.
  Phys.} {\bfseries 70} (2013) 159--209},
\href{http://arxiv.org/abs/1505.03828}{{\ttfamily arXiv:1505.03828 [nucl-ex]}}.

\bibitem{bib::Braun_Munziger}
P.~Braun-Munzinger, ``{Quarkonium production in ultra-relativistic nuclear
  collisions: Suppression versus enhancement},''
  \href{http://dx.doi.org/10.1088/0954-3899/34/8/S36}{{\em J. Phys.} {\bfseries
  G34} (2007) S471--478},
\href{http://arxiv.org/abs/nucl-th/0701093}{{\ttfamily arXiv:nucl-th/0701093
  [nucl-th]}}.

\bibitem{Colla}
M.~H. Thoma and M.~Gyulassy, ``Quark damping and energy loss in the high
  temperature {QCD},''
  \href{http://dx.doi.org/http://dx.doi.org/10.1016/S0550-3213(05)80031-8}{{\em
  Nuclear Physics} {\bfseries B351} no.~3, (1991) 491 -- 506}.
  \url{http://www.sciencedirect.com/science/article/pii/S0550321305800318}.

\bibitem{intro3}
M.~H. Thoma and M.~Gyulassy, ``{Quark Damping and Energy Loss in the High
  Temperature {QCD}},''
\href{http://dx.doi.org/10.1016/S0550-3213(05)80031-8}{{\em Nucl. Phys.}
  {\bfseries B351} (1991) 491--506}.

\bibitem{intro1}
M.~Gyulassy and M.~Plumer, ``{Jet Quenching in Dense Matter},''
\href{http://dx.doi.org/10.1016/0370-2693(90)91409-5}{{\em Phys. Lett.}
  {\bfseries B243} (1990) 432--438}.

\bibitem{bib::intro2}
R.~Baier, Y.~L. Dokshitzer, A.~H. Mueller, S.~Peigne, and D.~Schiff,
  ``{Radiative energy loss and $p_{\rm T}$-broadening of high-energy partons in
  nuclei},'' \href{http://dx.doi.org/10.1016/S0550-3213(96)00581-0}{{\em Nucl.
  Phys.} {\bfseries B484} (1997) 265--282},
\href{http://arxiv.org/abs/hep-ph/9608322}{{\ttfamily arXiv:hep-ph/9608322
  [hep-ph]}}.

\bibitem{bib::dead_cone_effect}
Y.~L. Dokshitzer and D.~E. Kharzeev, ``{Heavy quark colorimetry of QCD
  matter},'' \href{http://dx.doi.org/10.1016/S0370-2693(01)01130-3}{{\em Phys.
  Lett.} {\bfseries B519} (2001) 199--206},
\href{http://arxiv.org/abs/hep-ph/0106202}{{\ttfamily arXiv:hep-ph/0106202
  [hep-ph]}}.

\bibitem{bib::intro_massdep_Eloss1}
N.~Armesto, C.~A. Salgado, and U.~A. Wiedemann, ``{Medium induced gluon
  radiation off massive quarks fills the dead cone},''
  \href{http://dx.doi.org/10.1103/PhysRevD.69.114003}{{\em Phys. Rev.}
  {\bfseries D69} (2004) 114003},
\href{http://arxiv.org/abs/hep-ph/0312106}{{\ttfamily arXiv:hep-ph/0312106
  [hep-ph]}}.

\bibitem{bib::intro_massdep_Eloss2}
M.~Djordjevic and M.~Gyulassy, ``{Heavy quark radiative energy loss in QCD
  matter},'' \href{http://dx.doi.org/10.1016/j.nuclphysa.2003.12.020}{{\em
  Nucl. Phys.} {\bfseries A733} (2004) 265--298},
\href{http://arxiv.org/abs/nucl-th/0310076}{{\ttfamily arXiv:nucl-th/0310076
  [nucl-th]}}.

\bibitem{Zhang:2003wk}
B.-W. Zhang, E.~Wang, and X.-N. Wang, ``{Heavy quark energy loss in nuclear
  medium},'' \href{http://dx.doi.org/10.1103/PhysRevLett.93.072301}{{\em Phys.
  Rev. Lett.} {\bfseries 93} (2004) 072301},
\href{http://arxiv.org/abs/nucl-th/0309040}{{\ttfamily arXiv:nucl-th/0309040
  [nucl-th]}}.

\bibitem{bib::intro4}
H.~van Hees, V.~Greco, and R.~Rapp, ``{Heavy-quark probes of the quark-gluon
  plasma at RHIC},'' \href{http://dx.doi.org/10.1103/PhysRevC.73.034913}{{\em
  Phys. Rev.} {\bfseries C73} (2006) 034913},
\href{http://arxiv.org/abs/nucl-th/0508055}{{\ttfamily arXiv:nucl-th/0508055
  [nucl-th]}}.

\bibitem{Adam:2016wyz}
{\bfseries ALICE} Collaboration, J.~Adam {\em et~al.}, ``{Measurement of
  electrons from beauty-hadron decays in p-Pb collisions at $
  \sqrt{s_{\mathrm{NN}}}=5.02 $ TeV and Pb-Pb collisions at $
  \sqrt{s_{\mathrm{NN}}}=2.76 $ TeV},''
  \href{http://dx.doi.org/10.1007/JHEP07(2017)052}{{\em JHEP} {\bfseries 07}
  (2017) 052},
\href{http://arxiv.org/abs/1609.03898}{{\ttfamily arXiv:1609.03898 [nucl-ex]}}.

\bibitem{Adam:2015nna}
{\bfseries ALICE} Collaboration, J.~Adam {\em et~al.}, ``{Centrality dependence
  of high-p$_{T}$ D meson suppression in Pb-Pb collisions at $
  \sqrt{s_{\mathrm{N}\mathrm{N}}}=2.76 $ TeV},''
  \href{http://dx.doi.org/10.1007/JHEP11(2015)205,
  10.1007/JHEP06(2017)032}{{\em JHEP} {\bfseries 11} (2015) 205},
  \href{http://arxiv.org/abs/1506.06604}{{\ttfamily arXiv:1506.06604
  [nucl-ex]}}.
[Addendum: JHEP06,032(2017)].

\bibitem{Acharya:2018hre}
{\bfseries ALICE} Collaboration, S.~Acharya {\em et~al.}, ``{Measurement of
  D$^{0}$, D$^{+}$, D$^{*+}$ and D$_{s}^{+}$ production in Pb-Pb collisions at
  $ \sqrt{{\mathrm{s}}_{\mathrm{NN}}}=5.02 $ TeV},''
  \href{http://dx.doi.org/10.1007/JHEP10(2018)174}{{\em JHEP} {\bfseries 10}
  (2018) 174},
\href{http://arxiv.org/abs/1804.09083}{{\ttfamily arXiv:1804.09083 [nucl-ex]}}.

\bibitem{Sirunyan:2017oug}
{\bfseries CMS} Collaboration, A.~M. Sirunyan {\em et~al.}, ``{Measurement of
  the ${B}^{\pm}$ Meson Nuclear Modification Factor in Pb-Pb Collisions at
  $\sqrt{{s}_{\rm NN}}=5.02\mathrm{ TeV}$},''
  \href{http://dx.doi.org/10.1103/PhysRevLett.119.152301}{{\em Phys. Rev.
  Lett.} {\bfseries 119} no.~15, (2017) 152301},
\href{http://arxiv.org/abs/1705.04727}{{\ttfamily arXiv:1705.04727 [hep-ex]}}.

\bibitem{Adam:2015rba}
{\bfseries ALICE} Collaboration, J.~Adam {\em et~al.}, ``{Inclusive, prompt and
  non-prompt J/$\psi$ production at mid-rapidity in Pb-Pb collisions at
  $\sqrt{s_{\rm NN}}$ = 2.76 TeV},''
  \href{http://dx.doi.org/10.1007/JHEP07(2015)051}{{\em JHEP} {\bfseries 07}
  (2015) 051},
\href{http://arxiv.org/abs/1504.07151}{{\ttfamily arXiv:1504.07151 [nucl-ex]}}.

\bibitem{Sirunyan:2017isk}
{\bfseries CMS} Collaboration, A.~M. Sirunyan {\em et~al.}, ``{Measurement of
  prompt and nonprompt charmonium suppression in $\text {PbPb}$ collisions at
  5.02 $\,\text {Te}\text {V}$},''
  \href{http://dx.doi.org/10.1140/epjc/s10052-018-5950-6}{{\em Eur. Phys. J.}
  {\bfseries C78} no.~6, (2018) 509},
\href{http://arxiv.org/abs/1712.08959}{{\ttfamily arXiv:1712.08959 [nucl-ex]}}.

\bibitem{Sirunyan:2018ktu}
{\bfseries CMS} Collaboration, A.~M. Sirunyan {\em et~al.}, ``{Studies of
  Beauty Suppression via Nonprompt $D^0$ Mesons in Pb-Pb Collisions at $Q^2 =
  4$ $\rm GeV^2$},''
  \href{http://dx.doi.org/10.1103/PhysRevLett.123.022001}{{\em Phys. Rev.
  Lett.} {\bfseries 123} no.~2, (2019) 022001},
\href{http://arxiv.org/abs/1810.11102}{{\ttfamily arXiv:1810.11102 [hep-ex]}}.

\bibitem{TheATLAScollaboration}
{\bfseries ATLAS} Collaboration, M.~Aaboud {\em et~al.}, ``{Measurement of the
  suppression and azimuthal anisotropy of muons from heavy-flavor decays in
  Pb+Pb collisions at $\sqrt{s_{\mathrm{NN}}} = 2.76$ TeV with the ATLAS
  detector},'' \href{http://dx.doi.org/10.1103/PhysRevC.98.044905}{{\em Phys.
  Rev.} {\bfseries C98} no.~4, (2018) 044905},
\href{http://arxiv.org/abs/1805.05220}{{\ttfamily arXiv:1805.05220 [nucl-ex]}}.

\bibitem{Andronic:2015wma}
A.~Andronic {\em et~al.}, ``{Heavy-flavour and quarkonium production in the LHC
  era: from proton–proton to heavy-ion collisions},''
  \href{http://dx.doi.org/10.1140/epjc/s10052-015-3819-5}{{\em Eur. Phys. J.}
  {\bfseries C76} no.~3, (2016) 107},
\href{http://arxiv.org/abs/1506.03981}{{\ttfamily arXiv:1506.03981 [nucl-ex]}}.

\bibitem{Glauber:1970jm}
R.~J. Glauber and G.~Matthiae, ``{High-energy scattering of protons by
  nuclei},''
\href{http://dx.doi.org/10.1016/0550-3213(70)90511-0}{{\em Nucl. Phys.}
  {\bfseries B21} (1970) 135--157}.

\bibitem{Miller:2007ri}
M.~L. Miller, K.~Reygers, S.~J. Sanders, and P.~Steinberg, ``{Glauber modeling
  in high energy nuclear collisions},''
  \href{http://dx.doi.org/10.1146/annurev.nucl.57.090506.123020}{{\em Ann. Rev.
  Nucl. Part. Sci.} {\bfseries 57} (2007) 205--243},
\href{http://arxiv.org/abs/nucl-ex/0701025}{{\ttfamily arXiv:nucl-ex/0701025
  [nucl-ex]}}.

\bibitem{Abelev:2006db}
{\bfseries STAR} Collaboration, B.~I. Abelev {\em et~al.}, ``{Transverse
  momentum and centrality dependence of high-$p_T$ non-photonic electron
  suppression in Au+Au collisions at $\sqrt{s_{NN}} = 200$\,GeV},''
  \href{http://dx.doi.org/10.1103/PhysRevLett.106.159902,
  10.1103/PhysRevLett.98.192301}{{\em Phys. Rev. Lett.} {\bfseries 98} (2007)
  192301}, \href{http://arxiv.org/abs/nucl-ex/0607012}{{\ttfamily
  arXiv:nucl-ex/0607012 [nucl-ex]}}.
[Erratum: Phys. Rev. Lett.106,159902(2011)].

\bibitem{Adare:2010de}
{\bfseries PHENIX} Collaboration, A.~Adare {\em et~al.}, ``{Heavy Quark
  Production in $p+p$ and Energy Loss and Flow of Heavy Quarks in Au+Au
  Collisions at $\sqrt{s_{NN}}=200$ GeV},''
  \href{http://dx.doi.org/10.1103/PhysRevC.84.044905}{{\em Phys. Rev.}
  {\bfseries C84} (2011) 044905},
\href{http://arxiv.org/abs/1005.1627}{{\ttfamily arXiv:1005.1627 [nucl-ex]}}.

\bibitem{Adamczyk:2014uip}
{\bfseries STAR} Collaboration, L.~Adamczyk {\em et~al.}, ``{Observation of
  $D^0$ Meson Nuclear Modifications in Au+Au Collisions at $\sqrt{s_{NN}}=200$
  GeV},'' \href{http://dx.doi.org/10.1103/PhysRevLett.121.229901,
  10.1103/PhysRevLett.113.142301}{{\em Phys. Rev. Lett.} {\bfseries 113}
  no.~14, (2014) 142301}, \href{http://arxiv.org/abs/1404.6185}{{\ttfamily
  arXiv:1404.6185 [nucl-ex]}}.
[Erratum: Phys. Rev. Lett.121,no.22,229901(2018)].

\bibitem{Acharya:2018upq}
{\bfseries ALICE} Collaboration, S.~Acharya {\em et~al.}, ``{Measurements of
  low-p$_{T}$ electrons from semileptonic heavy-flavour hadron decays at
  mid-rapidity in pp and Pb-Pb collisions at $ \sqrt{s_{\mathrm{NN}}}=2.76 $
  TeV},'' \href{http://dx.doi.org/10.1007/JHEP10(2018)061}{{\em JHEP}
  {\bfseries 10} (2018) 061},
\href{http://arxiv.org/abs/1805.04379}{{\ttfamily arXiv:1805.04379 [nucl-ex]}}.

\bibitem{Adam:2016khe}
{\bfseries ALICE} Collaboration, J.~Adam {\em et~al.}, ``{Measurement of the
  production of high-$p_{\rm T}$ electrons from heavy-flavour hadron decays in
  Pb-Pb collisions at $\mathbf{\sqrt{\it s_{\rm{NN}}}}$ = 2.76 TeV},''
  \href{http://dx.doi.org/10.1016/j.physletb.2017.05.060}{{\em Phys. Lett.}
  {\bfseries B771} (2017) 467--481},
\href{http://arxiv.org/abs/1609.07104}{{\ttfamily arXiv:1609.07104 [nucl-ex]}}.

\bibitem{Adam:2016ssk}
{\bfseries ALICE} Collaboration, J.~Adam {\em et~al.}, ``{Elliptic flow of
  electrons from heavy-flavour hadron decays at mid-rapidity in Pb-Pb
  collisions at $\sqrt{{\mathrm{s}}_{\mathrm{NN}}}=2.76$ TeV},''
  \href{http://dx.doi.org/10.1007/JHEP09(2016)028}{{\em JHEP} {\bfseries 09}
  (2016) 028},
\href{http://arxiv.org/abs/1606.00321}{{\ttfamily arXiv:1606.00321 [nucl-ex]}}.

\bibitem{Abelev:2013lca}
{\bfseries ALICE} Collaboration, B.~Abelev {\em et~al.}, ``{{D} meson elliptic
  flow in non-central {{Pb--Pb}} collisions at energy $\sqrt{s_\mathrm{NN}}$ =
  2.76{TeV}},'' \href{http://dx.doi.org/10.1103/PhysRevLett.111.102301}{{\em
  Phys. Rev. Lett.} {\bfseries 111} (2013) 102301},
\href{http://arxiv.org/abs/1305.2707}{{\ttfamily arXiv:1305.2707 [nucl-ex]}}.

\bibitem{vanHees:2005wb}
H.~van Hees, V.~Greco, and R.~Rapp, ``{Heavy-quark probes of the quark-gluon
  plasma at RHIC},'' \href{http://dx.doi.org/10.1103/PhysRevC.73.034913}{{\em
  Phys. Rev.} {\bfseries C73} (2006) 034913},
\href{http://arxiv.org/abs/nucl-th/0508055}{{\ttfamily arXiv:nucl-th/0508055
  [nucl-th]}}.

\bibitem{Greco:2003vf}
V.~Greco, C.~M. Ko, and R.~Rapp, ``{Quark coalescence for charmed mesons in
  ultrarelativistic heavy ion collisions},''
  \href{http://dx.doi.org/10.1016/j.physletb.2004.06.064}{{\em Phys. Lett.}
  {\bfseries B595} (2004) 202--208},
\href{http://arxiv.org/abs/nucl-th/0312100}{{\ttfamily arXiv:nucl-th/0312100
  [nucl-th]}}.

\bibitem{Andronic:2003zv}
A.~Andronic, P.~Braun-Munzinger, K.~Redlich, and J.~Stachel, ``{Statistical
  hadronization of charm in heavy ion collisions at SPS, RHIC and LHC},''
  \href{http://dx.doi.org/10.1016/j.physletb.2003.07.066}{{\em Phys. Lett.}
  {\bfseries B571} (2003) 36--44},
\href{http://arxiv.org/abs/nucl-th/0303036}{{\ttfamily arXiv:nucl-th/0303036
  [nucl-th]}}.

\bibitem{Eskola:2009uj}
K.~J. Eskola, H.~Paukkunen, and C.~A. Salgado, ``{EPS09: A New Generation of
  NLO and LO Nuclear Parton Distribution Functions},''
  \href{http://dx.doi.org/10.1088/1126-6708/2009/04/065}{{\em JHEP} {\bfseries
  04} (2009) 065},
\href{http://arxiv.org/abs/0902.4154}{{\ttfamily arXiv:0902.4154 [hep-ph]}}.

\bibitem{Adam:2015qda}
{\bfseries ALICE} Collaboration, J.~Adam {\em et~al.}, ``{Measurement of
  electrons from heavy-flavour hadron decays in p-Pb collisions at
  $\sqrt{s_{\rm NN}} =$ 5.02 TeV},''
  \href{http://dx.doi.org/10.1016/j.physletb.2015.12.067}{{\em Phys. Lett.}
  {\bfseries B754} (2016) 81--93},
\href{http://arxiv.org/abs/1509.07491}{{\ttfamily arXiv:1509.07491 [nucl-ex]}}.

\bibitem{Abelev:2014hha}
{\bfseries ALICE} Collaboration, B.~Abelev {\em et~al.}, ``{Measurement of
  prompt $D$-meson production in $p-Pb$ collisions at $\sqrt{s_{\rm NN}}$ =
  5.02 TeV},'' \href{http://dx.doi.org/10.1103/PhysRevLett.113.232301}{{\em
  Phys. Rev. Lett.} {\bfseries 113} no.~23, (2014) 232301},
\href{http://arxiv.org/abs/1405.3452}{{\ttfamily arXiv:1405.3452 [nucl-ex]}}.

\bibitem{Acharya:2018dxy}
{\bfseries ALICE} Collaboration, S.~Acharya {\em et~al.}, ``{Azimuthal
  Anisotropy of Heavy-Flavor Decay Electrons in $p$-Pb Collisions at $
  \sqrt{s_{\rm NN}}$ = 5.02 TeV},''
  \href{http://dx.doi.org/10.1103/PhysRevLett.122.072301}{{\em Phys. Rev.
  Lett.} {\bfseries 122} no.~7, (2019) 072301},
\href{http://arxiv.org/abs/1805.04367}{{\ttfamily arXiv:1805.04367 [nucl-ex]}}.

\bibitem{Sirunyan:2018toe}
{\bfseries CMS} Collaboration, A.~M. Sirunyan {\em et~al.}, ``{Elliptic flow of
  charm and strange hadrons in high-multiplicity pPb collisions at
  $\sqrt{s_{_\mathrm{NN}}} =$ 8.16 TeV},''
  \href{http://dx.doi.org/10.1103/PhysRevLett.121.082301}{{\em Phys. Rev.
  Lett.} {\bfseries 121} no.~8, (2018) 082301},
\href{http://arxiv.org/abs/1804.09767}{{\ttfamily arXiv:1804.09767 [hep-ex]}}.

\bibitem{Aamodt:2008zz}
{\bfseries ALICE} Collaboration, K.~Aamodt {\em et~al.}, ``{The ALICE
  experiment at the CERN LHC},''
\href{http://dx.doi.org/10.1088/1748-0221/3/08/S08002}{{\em JINST} {\bfseries
  3} (2008) S08002}.

\bibitem{1748-0221-5-03-P03003}
{\bfseries ALICE} Collaboration, K.~Aamodt {\em et~al.}, ``{Alignment of the
  ALICE Inner Tracking System with cosmic-ray tracks},''
  \href{http://dx.doi.org/10.1088/1748-0221/5/03/P03003}{{\em JINST} {\bfseries
  5} (2010) P03003},
\href{http://arxiv.org/abs/1001.0502}{{\ttfamily arXiv:1001.0502
  [physics.ins-det]}}.

\bibitem{Alme:2010ke}
{\bfseries ALICE} Collaboration, J.~Alme, Y.~Andres, H.~Appelshauser,
  S.~Bablok, N.~Bialas, {\em et~al.}, ``{The {ALICE} {TPC}, a large
  3-dimensional tracking device with fast readout for ultra-high multiplicity
  events},'' \href{http://dx.doi.org/10.1016/j.nima.2010.04.042}{{\em
  Nucl.Instrum.Meth.} {\bfseries A622} (2010) 316--367},
\href{http://arxiv.org/abs/1001.1950}{{\ttfamily arXiv:1001.1950
  [physics.ins-det]}}.

\bibitem{Carnesecchi:2018oss}
{\bfseries ALICE} Collaboration, F.~Carnesecchi, ``{Performance of the ALICE
  Time-Of-Flight detector at the LHC},''
  \href{http://dx.doi.org/10.1088/1748-0221/14/06/C06023}{{\em JINST}
  {\bfseries 14} no.~06, (2019) C06023},
\href{http://arxiv.org/abs/1806.03825}{{\ttfamily arXiv:1806.03825
  [physics.ins-det]}}.

\bibitem{Abeysekara:2010ze}
{\bfseries ALICE EMCal} Collaboration, U.~Abeysekara {\em et~al.}, ``{ALICE
  EMCal Physics Performance Report},''
\href{http://arxiv.org/abs/1008.0413}{{\ttfamily arXiv:1008.0413
  [physics.ins-det]}}.

\bibitem{vZero}
{\bfseries ALICE} Collaboration, E.~Abbas {\em et~al.}, ``{Performance of the
  ALICE VZERO system},''
  \href{http://dx.doi.org/10.1088/1748-0221/8/10/P10016}{{\em JINST} {\bfseries
  8} (2013) P10016},
\href{http://arxiv.org/abs/1306.3130}{{\ttfamily arXiv:1306.3130 [nucl-ex]}}.

\bibitem{ALICE-PUBLIC-2018-011}
{\bfseries ALICE} Collaboration, ``{Centrality determination in heavy ion
  collisions},'' {\em
  {\href{https://cds.cern.ch/record/2636623?ln=en}{ALICE-PUBLIC-2018-011}}}
  (2018) . \url{https://cds.cern.ch/record/2636623}.

\bibitem{Loizides:2017ack}
C.~Loizides, J.~Kamin, and D.~d'Enterria, ``{Improved Monte Carlo Glauber
  predictions at present and future nuclear colliders},''
  \href{http://dx.doi.org/10.1103/PhysRevC.97.054910,
  10.1103/PhysRevC.99.019901}{{\em Phys. Rev.} {\bfseries C97} no.~5, (2018)
  054910}, \href{http://arxiv.org/abs/1710.07098}{{\ttfamily arXiv:1710.07098
  [nucl-ex]}}.
[erratum: Phys. Rev.C99,no.1,019901(2019)].

\bibitem{BetheBloch}
H.~Bethe, ``{Zur Theorie des Durchgangs schneller Korpuskularstrahlen durch
  Materie},'' \href{http://dx.doi.org/10.1002/andp.19303970303}{{\em Annalen
  der Physik} {\bfseries 397} no.~3, (1930) 325--400}.

\bibitem{Abelev:2014gla}
{\bfseries ALICE} Collaboration, B.~Abelev {\em et~al.}, ``{Measurement of
  electrons from semileptonic heavy-flavor hadron decays in $pp$ collisions at
  $\sqrt{s} = 2.76$ TeV},''
  \href{http://dx.doi.org/10.1103/PhysRevD.91.012001}{{\em Phys. Rev.}
  {\bfseries D91} no.~1, (2015) 012001},
\href{http://arxiv.org/abs/1405.4117}{{\ttfamily arXiv:1405.4117 [nucl-ex]}}.

\bibitem{Abelev:2012xe}
{\bfseries ALICE} Collaboration, B.~Abelev {\em et~al.}, ``{Measurement of
  electrons from semileptonic heavy-flavour hadron decays in pp collisions at
  $\sqrt{s}$ = 7 TeV},''
  \href{http://dx.doi.org/10.1103/PhysRevD.86.112007}{{\em Phys. Rev.}
  {\bfseries D86} (2012) 112007},
\href{http://arxiv.org/abs/1205.5423}{{\ttfamily arXiv:1205.5423 [hep-ex]}}.

\bibitem{Sjostrand:2006za}
T.~Sjostrand, S.~Mrenna, and P.~Z. Skands, ``{PYTHIA 6.4 Physics and Manual},''
  \href{http://dx.doi.org/10.1088/1126-6708/2006/05/026}{{\em JHEP} {\bfseries
  05} (2006) 026},
\href{http://arxiv.org/abs/hep-ph/0603175}{{\ttfamily arXiv:hep-ph/0603175
  [hep-ph]}}.

\bibitem{Hijing:ref}
M.~Gyulassy and X.-N. Wang, ``{HIJING 1.0: A Monte Carlo program for parton and
  particle production in high-energy hadronic and nuclear collisions},''
  \href{http://dx.doi.org/10.1016/0010-4655(94)90057-4}{{\em Comput. Phys.
  Commun.} {\bfseries 83} (1994) 307},
\href{http://arxiv.org/abs/nucl-th/9502021}{{\ttfamily arXiv:nucl-th/9502021
  [nucl-th]}}.

\bibitem{Brun:1082634}
R.~Brun, F.~Bruyant, F.~Carminati, S.~Giani, M.~Maire, A.~McPherson,
  G.~Patrick, and L.~Urban, {\em {GEANT: Detector Description and Simulation
  Tool; Oct 1994}}.
\newblock CERN Program Library. CERN, Geneva, 1993.
\newblock \url{https://cds.cern.ch/record/1082634}.
\newblock Long Writeup W5013.

\bibitem{collaboration2019production}
{\bfseries ALICE} Collaboration, S.~Acharya {\em et~al.}, ``{Production of
  charged pions, kaons and (anti-)protons in Pb-Pb and inelastic pp collisions
  at $\sqrt{s_{\rm{NN}}}$ = 5.02 TeV},''
  \href{http://arxiv.org/abs/1910.07678}{{\ttfamily arXiv:1910.07678
  [nucl-ex]}}.

\bibitem{Khandai:2011cf}
P.~K. Khandai, P.~Shukla, and V.~Singh, ``{Meson spectra and $m_T$ scaling in
  $p + p$, $d + $Au, and Au + Au collisions at $\sqrt{s_{NN}}=200$ GeV},''
  \href{http://dx.doi.org/10.1103/PhysRevC.84.054904}{{\em Phys. Rev.}
  {\bfseries C84} (2011) 054904},
\href{http://arxiv.org/abs/1110.3929}{{\ttfamily arXiv:1110.3929 [hep-ph]}}.

\bibitem{Altenkamper:2017qot}
L.~Altenkämper, F.~Bock, C.~Loizides, and N.~Schmidt, ``{Applicability of
  transverse mass scaling in hadronic collisions at energies available at the
  CERN Large Hadron Collider},''
  \href{http://dx.doi.org/10.1103/PhysRevC.96.064907}{{\em Phys. Rev.}
  {\bfseries C96} no.~6, (2017) 064907},
\href{http://arxiv.org/abs/1710.01933}{{\ttfamily arXiv:1710.01933 [hep-ph]}}.

\bibitem{Oleari:2010nx}
C.~Oleari, ``{The POWHEG-BOX},''
  \href{http://dx.doi.org/10.1016/j.nuclphysbps.2010.08.016}{{\em Nucl. Phys.
  Proc. Suppl.} {\bfseries 205-206} (2010) 36--41},
\href{http://arxiv.org/abs/1007.3893}{{\ttfamily arXiv:1007.3893 [hep-ph]}}.

\bibitem{ALICE-PUBLIC-2018-014}
{\bfseries ALICE} Collaboration, ``{ALICE 2017 luminosity determination for pp
  collisions at $\sqrt{s}$ = 5 TeV},'' {\em
  {\href{http://cds.cern.ch/record/2648933}{ALICE-PUBLIC-2018-014}}} (Nov,
  2018) . \url{http://cds.cern.ch/record/2648933}.

\bibitem{Aad:2014qxa}
{\bfseries ATLAS} Collaboration, G.~Aad {\em et~al.}, ``{Measurements of the W
  production cross sections in association with jets with the ATLAS
  detector},'' \href{http://dx.doi.org/10.1140/epjc/s10052-015-3262-7}{{\em
  Eur. Phys. J.} {\bfseries C75} no.~2, (2015) 82},
\href{http://arxiv.org/abs/1409.8639}{{\ttfamily arXiv:1409.8639 [hep-ex]}}.

\bibitem{Aad:2014qja}
{\bfseries ATLAS} Collaboration, G.~Aad {\em et~al.}, ``{Measurement of the
  low-mass Drell-Yan differential cross section at $\sqrt{s}$ = 7 TeV using the
  ATLAS detector},'' \href{http://dx.doi.org/10.1007/JHEP06(2014)112}{{\em
  JHEP} {\bfseries 06} (2014) 112},
\href{http://arxiv.org/abs/1404.1212}{{\ttfamily arXiv:1404.1212 [hep-ex]}}.

\bibitem{Sirunyan:2018owv}
{\bfseries CMS} Collaboration, A.~M. Sirunyan {\em et~al.}, ``{Measurement of
  the differential Drell-Yan cross section in proton-proton collisions at $
  \sqrt{\mathrm{s}} $ = 13 TeV},''
  \href{http://dx.doi.org/10.1007/JHEP12(2019)059}{{\em JHEP} {\bfseries 12}
  (2019) 059},
\href{http://arxiv.org/abs/1812.10529}{{\ttfamily arXiv:1812.10529 [hep-ex]}}.

\bibitem{Adam:2015sza}
{\bfseries ALICE} Collaboration, J.~Adam {\em et~al.}, ``{Transverse momentum
  dependence of D-meson production in Pb-Pb collisions at $
  \sqrt{{\mathrm{s}}_{\mathrm{NN}}}=$ 2.76 TeV},''
  \href{http://dx.doi.org/10.1007/JHEP03(2016)081}{{\em JHEP} {\bfseries 03}
  (2016) 081},
\href{http://arxiv.org/abs/1509.06888}{{\ttfamily arXiv:1509.06888 [nucl-ex]}}.

\bibitem{Cacciari:1998it}
M.~Cacciari, M.~Greco, and P.~Nason, ``{The $p_T$ spectrum in heavy flavor
  hadroproduction},''
  \href{http://dx.doi.org/10.1088/1126-6708/1998/05/007}{{\em JHEP} {\bfseries
  05} (1998) 007},
\href{http://arxiv.org/abs/hep-ph/9803400}{{\ttfamily arXiv:hep-ph/9803400
  [hep-ph]}}.

\bibitem{Nadolsky:2008zw}
P.~M. Nadolsky, H.-L. Lai, Q.-H. Cao, J.~Huston, J.~Pumplin, D.~Stump, W.-K.
  Tung, and C.~P. Yuan, ``{Implications of CTEQ global analysis for collider
  observables},'' \href{http://dx.doi.org/10.1103/PhysRevD.78.013004}{{\em
  Phys. Rev.} {\bfseries D78} (2008) 013004},
\href{http://arxiv.org/abs/0802.0007}{{\ttfamily arXiv:0802.0007 [hep-ph]}}.

\bibitem{Aad:2011rr}
{\bfseries ATLAS} Collaboration, G.~Aad {\em et~al.}, ``{Measurements of the
  electron and muon inclusive cross-sections in proton-proton collisions at
  $\sqrt{s}=7$ TeV with the ATLAS detector},''
  \href{http://dx.doi.org/10.1016/j.physletb.2011.12.054}{{\em Phys. Lett.}
  {\bfseries B707} (2012) 438--458},
\href{http://arxiv.org/abs/1109.0525}{{\ttfamily arXiv:1109.0525 [hep-ex]}}.

\bibitem{Averbeck:2011ga}
R.~Averbeck, N.~Bastid, Z.~C. del Valle, P.~Crochet, A.~Dainese, and X.~Zhang,
  ``{Reference Heavy Flavour Cross Sections in pp Collisions at
  $\sqrt{\mathrm{s}}$ = 2.76 TeV, using a pQCD-Driven
  $\sqrt{\mathrm{s}}$-Scaling of ALICE Measurements at $\sqrt{\mathrm{s}}$ = 7
  TeV},''
\href{http://arxiv.org/abs/1107.3243}{{\ttfamily arXiv:1107.3243 [hep-ph]}}.

\bibitem{Acharya:2018njl}
{\bfseries ALICE} Collaboration, S.~Acharya {\em et~al.}, ``{Analysis of the
  apparent nuclear modification in peripheral Pb–Pb collisions at 5.02
  TeV},'' \href{http://dx.doi.org/10.1016/j.physletb.2019.04.047}{{\em Phys.
  Lett.} {\bfseries B793} (2019) 420--432},
\href{http://arxiv.org/abs/1805.05212}{{\ttfamily arXiv:1805.05212 [nucl-ex]}}.

\bibitem{Morsch:2017brb}
C.~Loizides and A.~Morsch, ``{Absence of jet quenching in peripheral
  nucleus–nucleus collisions},''
  \href{http://dx.doi.org/10.1016/j.physletb.2017.09.002}{{\em Phys. Lett.}
  {\bfseries B773} (2017) 408--411},
\href{http://arxiv.org/abs/1705.08856}{{\ttfamily arXiv:1705.08856 [nucl-ex]}}.

\bibitem{Acharya:2018ckj}
{\bfseries ALICE} Collaboration, S.~Acharya {\em et~al.},
  ``{$\Lambda_\mathrm{c}^+$ production in Pb-Pb collisions at $\sqrt{s_{\rm
  NN}} = 5.02$ TeV},''
  \href{http://dx.doi.org/10.1016/j.physletb.2019.04.046}{{\em Phys. Lett.}
  {\bfseries B793} (2019) 212--223},
\href{http://arxiv.org/abs/1809.10922}{{\ttfamily arXiv:1809.10922 [nucl-ex]}}.

\bibitem{He:2011qa}
M.~He, R.~J. Fries, and R.~Rapp, ``{Heavy-Quark Diffusion and Hadronization in
  Quark-Gluon Plasma},''
  \href{http://dx.doi.org/10.1103/PhysRevC.86.014903}{{\em Phys. Rev.}
  {\bfseries C86} (2012) 014903},
\href{http://arxiv.org/abs/1106.6006}{{\ttfamily arXiv:1106.6006 [nucl-th]}}.

\bibitem{He:2012df}
M.~He, R.~J. Fries, and R.~Rapp, ``{$\mathbf{D_s}$-Meson as Quantitative Probe
  of Diffusion and Hadronization in Nuclear Collisions},''
  \href{http://dx.doi.org/10.1103/PhysRevLett.110.112301}{{\em Phys. Rev.
  Lett.} {\bfseries 110} no.~11, (2013) 112301},
\href{http://arxiv.org/abs/1204.4442}{{\ttfamily arXiv:1204.4442 [nucl-th]}}.

\bibitem{Gossiaux:2010yx}
P.~B. Gossiaux, J.~Aichelin, T.~Gousset, and V.~Guiho, ``{Competition of Heavy
  Quark Radiative and Collisional Energy Loss in Deconfined Matter},''
  \href{http://dx.doi.org/10.1088/0954-3899/37/9/094019}{{\em J. Phys.}
  {\bfseries G37} (2010) 094019},
\href{http://arxiv.org/abs/1001.4166}{{\ttfamily arXiv:1001.4166 [hep-ph]}}.

\bibitem{Gossiaux:2012ya}
P.~B. Gossiaux, M.~Nahrgang, M.~Bluhm, T.~Gousset, and J.~Aichelin, ``{Heavy
  quark quenching from RHIC to LHC and the consequences of gluon damping},''
  \href{http://dx.doi.org/10.1016/j.nuclphysa.2013.02.182}{{\em Nucl. Phys.}
  {\bfseries A904-905} (2013) 992c--995c},
\href{http://arxiv.org/abs/1211.2281}{{\ttfamily arXiv:1211.2281 [hep-ph]}}.

\bibitem{Djordjevic:2015hra}
M.~Djordjevic and M.~Djordjevic, ``{Predictions of heavy-flavor suppression at
  5.1 TeV Pb + Pb collisions at the CERN Large Hadron Collider},''
  \href{http://dx.doi.org/10.1103/PhysRevC.92.024918}{{\em Phys. Rev.}
  {\bfseries C92} no.~2, (2015) 024918},
\href{http://arxiv.org/abs/1505.04316}{{\ttfamily arXiv:1505.04316 [nucl-th]}}.

\bibitem{Xu:2015bbz}
J.~Xu, J.~Liao, and M.~Gyulassy, ``{Bridging Soft-Hard Transport Properties of
  Quark-Gluon Plasmas with CUJET3.0},''
  \href{http://dx.doi.org/10.1007/JHEP02(2016)169}{{\em JHEP} {\bfseries 02}
  (2016) 169},
\href{http://arxiv.org/abs/1508.00552}{{\ttfamily arXiv:1508.00552 [hep-ph]}}.

\bibitem{Uphoff:2014hza}
J.~Uphoff, O.~Fochler, Z.~Xu, and C.~Greiner, ``{Elastic and radiative heavy
  quark interactions in ultra-relativistic heavy-ion collisions},''
  \href{http://dx.doi.org/10.1088/0954-3899/42/11/115106}{{\em J. Phys.}
  {\bfseries G42} no.~11, (2015) 115106},
\href{http://arxiv.org/abs/1408.2964}{{\ttfamily arXiv:1408.2964 [hep-ph]}}.

\bibitem{Song:2015ykw}
T.~Song, H.~Berrehrah, D.~Cabrera, W.~Cassing, and E.~Bratkovskaya, ``{Charm
  production in Pb + Pb collisions at energies available at the CERN Large
  Hadron Collider},'' \href{http://dx.doi.org/10.1103/PhysRevC.93.034906}{{\em
  Phys. Rev.} {\bfseries C93} no.~3, (2016) 034906},
\href{http://arxiv.org/abs/1512.00891}{{\ttfamily arXiv:1512.00891 [nucl-th]}}.

\bibitem{He:2014cla}
M.~He, R.~J. Fries, and R.~Rapp, ``{Heavy Flavor at the Large Hadron Collider
  in a Strong Coupling Approach},''
  \href{http://dx.doi.org/10.1016/j.physletb.2014.05.050}{{\em Phys.Lett.}
  {\bfseries B735} (2014) 445--450},
\href{http://arxiv.org/abs/1401.3817}{{\ttfamily arXiv:1401.3817 [nucl-th]}}.

\bibitem{Djordjevic:2014}
M.~Djordjevic and M.~Djordjevic, ``{LHC jet suppression of light and heavy
  flavor observables},''
  \href{http://dx.doi.org/10.1016/j.physletb.2014.05.053}{{\em Phys. Lett.}
  {\bfseries B734} (2014) 286--289},
\href{http://arxiv.org/abs/1307.4098}{{\ttfamily arXiv:1307.4098 [hep-ph]}}.

\bibitem{Beraudo:2014boa}
A.~Beraudo, A.~De~Pace, M.~Monteno, M.~Nardi, and F.~Prino, ``{Heavy flavors in
  heavy-ion collisions: quenching, flow and correlations},''
  \href{http://dx.doi.org/10.1140/epjc/s10052-015-3336-6}{{\em Eur. Phys. J.}
  {\bfseries C75} no.~3, (2015) 121},
\href{http://arxiv.org/abs/1410.6082}{{\ttfamily arXiv:1410.6082 [hep-ph]}}.

\bibitem{Nahrgang:2013xaa}
M.~Nahrgang, J.~Aichelin, P.~B. Gossiaux, and K.~Werner, ``{Influence of
  hadronic bound states above $T_{\rm c}$ on heavy-quark observables in
  {{Pb--Pb}} collisions at at the CERN Large Hadron Collider},''
  \href{http://dx.doi.org/10.1103/PhysRevC.89.014905}{{\em Phys.Rev.}
  {\bfseries C89} no.~1, (2014) 014905},
\href{http://arxiv.org/abs/1305.6544}{{\ttfamily arXiv:1305.6544 [hep-ph]}}.

\end{thebibliography}\endgroup


\providecommand{\href}[2]{#2}\begingroup\raggedright\endgroup

\newpage
\appendix
\section{The ALICE Collaboration}
\label{app:collab}

\begingroup
\small
\begin{flushleft}
S.~Acharya\Irefn{org141}\And 
D.~Adamov\'{a}\Irefn{org93}\And 
S.P.~Adhya\Irefn{org141}\And 
A.~Adler\Irefn{org73}\And 
J.~Adolfsson\Irefn{org79}\And 
M.M.~Aggarwal\Irefn{org98}\And 
G.~Aglieri Rinella\Irefn{org34}\And 
M.~Agnello\Irefn{org31}\And 
N.~Agrawal\Irefn{org10}\textsuperscript{,}\Irefn{org48}\textsuperscript{,}\Irefn{org53}\And 
Z.~Ahammed\Irefn{org141}
% ...
\And
\renewcommand\labelenumi{\textsuperscript{\theenumi}~}

\section*{Affiliation notes}
\renewcommand\theenumi{\roman{enumi}}
\begin{Authlist}
\item \Adef{org*}Deceased
\item \Adef{orgI}Dipartimento DET del Politecnico di Torino, Turin, Italy
\item \Adef{orgII}M.V. Lomonosov Moscow State University, D.V. Skobeltsyn Institute of Nuclear, Physics, Moscow, Russia
\item \Adef{orgIII}Department of Applied Physics, Aligarh Muslim University, Aligarh, India
\item \Adef{orgIV}Institute of Theoretical Physics, University of Wroclaw, Poland
\end{Authlist}

\section*{Collaboration Institutes}
\renewcommand\theenumi{\arabic{enumi}~}
\begin{Authlist}
\item \Idef{org1}A.I. Alikhanyan National Science Laboratory (Yerevan Physics Institute) Foundation, Yerevan, Armenia
\item \Idef{org2}Bogolyubov Institute for Theoretical Physics, National Academy of Sciences of Ukraine, Kiev, Ukraine
\item \Idef{org3}Bose Institute, Department of Physics  and Centre for Astroparticle Physics and Space Science (CAPSS), Kolkata, India
\item \Idef{org4}Budker Institute for Nuclear Physics, Novosibirsk, Russia
\item \Idef{org5}California Polytechnic State University, San Luis Obispo, California, United States
\item \Idef{org6}Central China Normal University, Wuhan, China
\item \Idef{org7}Centre de Calcul de l'IN2P3, Villeurbanne, Lyon, France
\item \Idef{org8}Centro de Aplicaciones Tecnol\'{o}gicas y Desarrollo Nuclear (CEADEN), Havana, Cuba
\item \Idef{org9}Centro de Investigaci\'{o}n y de Estudios Avanzados (CINVESTAV), Mexico City and M\'{e}rida, Mexico
\item \Idef{org10}Centro Fermi - Museo Storico della Fisica e Centro Studi e Ricerche ``Enrico Fermi', Rome, Italy
\item \Idef{org11}Chicago State University, Chicago, Illinois, United States
\item \Idef{org12}China Institute of Atomic Energy, Beijing, China
\item \Idef{org13}Chonbuk National University, Jeonju, Republic of Korea
\item \Idef{org14}Comenius University Bratislava, Faculty of Mathematics, Physics and Informatics, Bratislava, Slovakia
\item \Idef{org15}COMSATS University Islamabad, Islamabad, Pakistan
\item \Idef{org16}Creighton University, Omaha, Nebraska, United States
\item \Idef{org17}Department of Physics, Aligarh Muslim University, Aligarh, India
\item \Idef{org18}Department of Physics, Pusan National University, Pusan, Republic of Korea
\item \Idef{org19}Department of Physics, Sejong University, Seoul, Republic of Korea
\item \Idef{org20}Department of Physics, University of California, Berkeley, California, United States
\item \Idef{org21}Department of Physics, University of Oslo, Oslo, Norway
\item \Idef{org22}Department of Physics and Technology, University of Bergen, Bergen, Norway
\item \Idef{org23}Dipartimento di Fisica dell'Universit\`{a} 'La Sapienza' and Sezione INFN, Rome, Italy
\item \Idef{org24}Dipartimento di Fisica dell'Universit\`{a} and Sezione INFN, Cagliari, Italy
\item \Idef{org25}Dipartimento di Fisica dell'Universit\`{a} and Sezione INFN, Trieste, Italy
\item \Idef{org26}Dipartimento di Fisica dell'Universit\`{a} and Sezione INFN, Turin, Italy
\item \Idef{org27}Dipartimento di Fisica e Astronomia dell'Universit\`{a} and Sezione INFN, Bologna, Italy
\item \Idef{org28}Dipartimento di Fisica e Astronomia dell'Universit\`{a} and Sezione INFN, Catania, Italy
\item \Idef{org29}Dipartimento di Fisica e Astronomia dell'Universit\`{a} and Sezione INFN, Padova, Italy
\item \Idef{org30}Dipartimento di Fisica `E.R.~Caianiello' dell'Universit\`{a} and Gruppo Collegato INFN, Salerno, Italy
\item \Idef{org31}Dipartimento DISAT del Politecnico and Sezione INFN, Turin, Italy
\item \Idef{org32}Dipartimento di Scienze e Innovazione Tecnologica dell'Universit\`{a} del Piemonte Orientale and INFN Sezione di Torino, Alessandria, Italy
\item \Idef{org33}Dipartimento Interateneo di Fisica `M.~Merlin' and Sezione INFN, Bari, Italy
\item \Idef{org34}European Organization for Nuclear Research (CERN), Geneva, Switzerland
\item \Idef{org35}Faculty of Electrical Engineering, Mechanical Engineering and Naval Architecture, University of Split, Split, Croatia
\item \Idef{org36}Faculty of Engineering and Science, Western Norway University of Applied Sciences, Bergen, Norway
\item \Idef{org37}Faculty of Nuclear Sciences and Physical Engineering, Czech Technical University in Prague, Prague, Czech Republic
\item \Idef{org38}Faculty of Science, P.J.~\v{S}af\'{a}rik University, Ko\v{s}ice, Slovakia
\item \Idef{org39}Frankfurt Institute for Advanced Studies, Johann Wolfgang Goethe-Universit\"{a}t Frankfurt, Frankfurt, Germany
\item \Idef{org40}Gangneung-Wonju National University, Gangneung, Republic of Korea
\item \Idef{org41}Gauhati University, Department of Physics, Guwahati, India
\item \Idef{org42}Helmholtz-Institut f\"{u}r Strahlen- und Kernphysik, Rheinische Friedrich-Wilhelms-Universit\"{a}t Bonn, Bonn, Germany
\item \Idef{org43}Helsinki Institute of Physics (HIP), Helsinki, Finland
\item \Idef{org44}High Energy Physics Group,  Universidad Aut\'{o}noma de Puebla, Puebla, Mexico
\item \Idef{org45}Hiroshima University, Hiroshima, Japan
\item \Idef{org46}Hochschule Worms, Zentrum  f\"{u}r Technologietransfer und Telekommunikation (ZTT), Worms, Germany
\item \Idef{org47}Horia Hulubei National Institute of Physics and Nuclear Engineering, Bucharest, Romania
\item \Idef{org48}Indian Institute of Technology Bombay (IIT), Mumbai, India
\item \Idef{org49}Indian Institute of Technology Indore, Indore, India
\item \Idef{org50}Indonesian Institute of Sciences, Jakarta, Indonesia
\item \Idef{org51}INFN, Laboratori Nazionali di Frascati, Frascati, Italy
\item \Idef{org52}INFN, Sezione di Bari, Bari, Italy
\item \Idef{org53}INFN, Sezione di Bologna, Bologna, Italy
\item \Idef{org54}INFN, Sezione di Cagliari, Cagliari, Italy
\item \Idef{org55}INFN, Sezione di Catania, Catania, Italy
\item \Idef{org56}INFN, Sezione di Padova, Padova, Italy
\item \Idef{org57}INFN, Sezione di Roma, Rome, Italy
\item \Idef{org58}INFN, Sezione di Torino, Turin, Italy
\item \Idef{org59}INFN, Sezione di Trieste, Trieste, Italy
\item \Idef{org60}Inha University, Incheon, Republic of Korea
\item \Idef{org61}Institut de Physique Nucl\'{e}aire d'Orsay (IPNO), Institut National de Physique Nucl\'{e}aire et de Physique des Particules (IN2P3/CNRS), Universit\'{e} de Paris-Sud, Universit\'{e} Paris-Saclay, Orsay, France
\item \Idef{org62}Institute for Nuclear Research, Academy of Sciences, Moscow, Russia
\item \Idef{org63}Institute for Subatomic Physics, Utrecht University/Nikhef, Utrecht, Netherlands
\item \Idef{org64}Institute of Experimental Physics, Slovak Academy of Sciences, Ko\v{s}ice, Slovakia
\item \Idef{org65}Institute of Physics, Homi Bhabha National Institute, Bhubaneswar, India
\item \Idef{org66}Institute of Physics of the Czech Academy of Sciences, Prague, Czech Republic
\item \Idef{org67}Institute of Space Science (ISS), Bucharest, Romania
\item \Idef{org68}Institut f\"{u}r Kernphysik, Johann Wolfgang Goethe-Universit\"{a}t Frankfurt, Frankfurt, Germany
\item \Idef{org69}Instituto de Ciencias Nucleares, Universidad Nacional Aut\'{o}noma de M\'{e}xico, Mexico City, Mexico
\item \Idef{org70}Instituto de F\'{i}sica, Universidade Federal do Rio Grande do Sul (UFRGS), Porto Alegre, Brazil
\item \Idef{org71}Instituto de F\'{\i}sica, Universidad Nacional Aut\'{o}noma de M\'{e}xico, Mexico City, Mexico
\item \Idef{org72}iThemba LABS, National Research Foundation, Somerset West, South Africa
\item \Idef{org73}Johann-Wolfgang-Goethe Universit\"{a}t Frankfurt Institut f\"{u}r Informatik, Fachbereich Informatik und Mathematik, Frankfurt, Germany
\item \Idef{org74}Joint Institute for Nuclear Research (JINR), Dubna, Russia
\item \Idef{org75}Korea Institute of Science and Technology Information, Daejeon, Republic of Korea
\item \Idef{org76}KTO Karatay University, Konya, Turkey
\item \Idef{org77}Laboratoire de Physique Subatomique et de Cosmologie, Universit\'{e} Grenoble-Alpes, CNRS-IN2P3, Grenoble, France
\item \Idef{org78}Lawrence Berkeley National Laboratory, Berkeley, California, United States
\item \Idef{org79}Lund University Department of Physics, Division of Particle Physics, Lund, Sweden
\item \Idef{org80}Nagasaki Institute of Applied Science, Nagasaki, Japan
\item \Idef{org81}Nara Women{'}s University (NWU), Nara, Japan
\item \Idef{org82}National and Kapodistrian University of Athens, School of Science, Department of Physics , Athens, Greece
\item \Idef{org83}National Centre for Nuclear Research, Warsaw, Poland
\item \Idef{org84}National Institute of Science Education and Research, Homi Bhabha National Institute, Jatni, India
\item \Idef{org85}National Nuclear Research Center, Baku, Azerbaijan
\item \Idef{org86}National Research Centre Kurchatov Institute, Moscow, Russia
\item \Idef{org87}Niels Bohr Institute, University of Copenhagen, Copenhagen, Denmark
\item \Idef{org88}Nikhef, National institute for subatomic physics, Amsterdam, Netherlands
\item \Idef{org89}NRC Kurchatov Institute IHEP, Protvino, Russia
\item \Idef{org90}NRC  « Kurchatov Institute »  - ITEP, Moscow, Russia
\item \Idef{org91}NRNU Moscow Engineering Physics Institute, Moscow, Russia
\item \Idef{org92}Nuclear Physics Group, STFC Daresbury Laboratory, Daresbury, United Kingdom
\item \Idef{org93}Nuclear Physics Institute of the Czech Academy of Sciences, \v{R}e\v{z} u Prahy, Czech Republic
\item \Idef{org94}Oak Ridge National Laboratory, Oak Ridge, Tennessee, United States
\item \Idef{org95}Ohio State University, Columbus, Ohio, United States
\item \Idef{org96}Petersburg Nuclear Physics Institute, Gatchina, Russia
\item \Idef{org97}Physics department, Faculty of science, University of Zagreb, Zagreb, Croatia
\item \Idef{org98}Physics Department, Panjab University, Chandigarh, India
\item \Idef{org99}Physics Department, University of Jammu, Jammu, India
\item \Idef{org100}Physics Department, University of Rajasthan, Jaipur, India
\item \Idef{org101}Physikalisches Institut, Eberhard-Karls-Universit\"{a}t T\"{u}bingen, T\"{u}bingen, Germany
\item \Idef{org102}Physikalisches Institut, Ruprecht-Karls-Universit\"{a}t Heidelberg, Heidelberg, Germany
\item \Idef{org103}Physik Department, Technische Universit\"{a}t M\"{u}nchen, Munich, Germany
\item \Idef{org104}Politecnico di Bari, Bari, Italy
\item \Idef{org105}Research Division and ExtreMe Matter Institute EMMI, GSI Helmholtzzentrum f\"ur Schwerionenforschung GmbH, Darmstadt, Germany
\item \Idef{org106}Rudjer Bo\v{s}kovi\'{c} Institute, Zagreb, Croatia
\item \Idef{org107}Russian Federal Nuclear Center (VNIIEF), Sarov, Russia
\item \Idef{org108}Saha Institute of Nuclear Physics, Homi Bhabha National Institute, Kolkata, India
\item \Idef{org109}School of Physics and Astronomy, University of Birmingham, Birmingham, United Kingdom
\item \Idef{org110}Secci\'{o}n F\'{\i}sica, Departamento de Ciencias, Pontificia Universidad Cat\'{o}lica del Per\'{u}, Lima, Peru
\item \Idef{org111}Shanghai Institute of Applied Physics, Shanghai, China
\item \Idef{org112}St. Petersburg State University, St. Petersburg, Russia
\item \Idef{org113}Stefan Meyer Institut f\"{u}r Subatomare Physik (SMI), Vienna, Austria
\item \Idef{org114}SUBATECH, IMT Atlantique, Universit\'{e} de Nantes, CNRS-IN2P3, Nantes, France
\item \Idef{org115}Suranaree University of Technology, Nakhon Ratchasima, Thailand
\item \Idef{org116}Technical University of Ko\v{s}ice, Ko\v{s}ice, Slovakia
\item \Idef{org117}Technische Universit\"{a}t M\"{u}nchen, Excellence Cluster 'Universe', Munich, Germany
\item \Idef{org118}The Henryk Niewodniczanski Institute of Nuclear Physics, Polish Academy of Sciences, Cracow, Poland
\item \Idef{org119}The University of Texas at Austin, Austin, Texas, United States
\item \Idef{org120}Universidad Aut\'{o}noma de Sinaloa, Culiac\'{a}n, Mexico
\item \Idef{org121}Universidade de S\~{a}o Paulo (USP), S\~{a}o Paulo, Brazil
\item \Idef{org122}Universidade Estadual de Campinas (UNICAMP), Campinas, Brazil
\item \Idef{org123}Universidade Federal do ABC, Santo Andre, Brazil
\item \Idef{org124}University of Cape Town, Cape Town, South Africa
\item \Idef{org125}University of Houston, Houston, Texas, United States
\item \Idef{org126}University of Jyv\"{a}skyl\"{a}, Jyv\"{a}skyl\"{a}, Finland
\item \Idef{org127}University of Liverpool, Liverpool, United Kingdom
\item \Idef{org128}University of Science and Technology of China, Hefei, China
\item \Idef{org129}University of South-Eastern Norway, Tonsberg, Norway
\item \Idef{org130}University of Tennessee, Knoxville, Tennessee, United States
\item \Idef{org131}University of the Witwatersrand, Johannesburg, South Africa
\item \Idef{org132}University of Tokyo, Tokyo, Japan
\item \Idef{org133}University of Tsukuba, Tsukuba, Japan
\item \Idef{org134}Universit\'{e} Clermont Auvergne, CNRS/IN2P3, LPC, Clermont-Ferrand, France
\item \Idef{org135}Universit\'{e} de Lyon, Universit\'{e} Lyon 1, CNRS/IN2P3, IPN-Lyon, Villeurbanne, Lyon, France
\item \Idef{org136}Universit\'{e} de Strasbourg, CNRS, IPHC UMR 7178, F-67000 Strasbourg, France, Strasbourg, France
\item \Idef{org137}Universit\'{e} Paris-Saclay Centre d'Etudes de Saclay (CEA), IRFU, D\'{e}partment de Physique Nucl\'{e}aire (DPhN), Saclay, France
\item \Idef{org138}Universit\`{a} degli Studi di Foggia, Foggia, Italy
\item \Idef{org139}Universit\`{a} degli Studi di Pavia, Pavia, Italy
\item \Idef{org140}Universit\`{a} di Brescia, Brescia, Italy
\item \Idef{org141}Variable Energy Cyclotron Centre, Homi Bhabha National Institute, Kolkata, India
\item \Idef{org142}Warsaw University of Technology, Warsaw, Poland
\item \Idef{org143}Wayne State University, Detroit, Michigan, United States
\item \Idef{org144}Westf\"{a}lische Wilhelms-Universit\"{a}t M\"{u}nster, Institut f\"{u}r Kernphysik, M\"{u}nster, Germany
\item \Idef{org145}Wigner Research Centre for Physics, Hungarian Academy of Sciences, Budapest, Hungary
\item \Idef{org146}Yale University, New Haven, Connecticut, United States
\item \Idef{org147}Yonsei University, Seoul, Republic of Korea
\end{Authlist}
\endgroup
\end{document}